\documentclass[pre,aps,showpacs,twocolumn]{revtex4}
\usepackage[dvips]{graphicx}
\newtheorem{definition}{Definition}

\begin{document}

\title{Using the fractional interaction law to model the impact dynamics in
       arbitrary form of multiparticle collisions}
\author{Jacek S. Leszczynski}
\email{jale@k2.pcz.czest.pl}
\affiliation{Czestochowa University of Technology, Institute of Mathematics and
             Computer Science, ul. Dabrowskiego 73, 42-200 Czestochowa, Poland}

\begin{abstract}
Using the molecular dynamics method, we examine a~discrete deterministic model
for the motion of spherical particles in three-dimensional space. The model takes
into account multiparticle collisions in arbitrary forms. Using fractional
calculus we proposed an expression for the repulsive force, which is the so
called fractional interaction law. We then illustrate and discuss how to
control (correlate) the energy dissipation and the collisional time for an
individual particle within multiparticle collisions. In the multiparticle
collisions we included the friction mechanism needed for the transition from
coupled torsion-sliding friction through rolling friction to static friction.
Analysing simple simulations we found that in the strong repulsive state binary
collisions dominate. However, within multiparticle collisions weak repulsion
is observed to be much stronger. The presented numerical results can be used to
realistically model the impact dynamics of an individual particle in a~group
of colliding particles.
\end{abstract}

\maketitle

\section{Introduction}
Nature of particulate flows offers the physics and engineering communities an
opportunity to analyse the interesting behaviour of granular materials. From
a~phenomenological point of view such a~flow, being halfway between a~solid and
a~liquid state, is not well understood because the basic physics is extremely
complex. In a~local state, the simplest form of the granular dynamics is as
follows - during an arbitrary extortion particles move individually and during
collisions particles may exchange their momenta and energies. Therefore the
collision process plays a~dominant role in the development of theoretical
studies and also in the performance of simulations. For an understanding of the
collision process we need to consider a~simple situation, focusing on what
happens when two particles collide. In other words, we need to be able to
distinguish the following basic phenomena: static contact~\cite{Geng},
cohesion~\cite{Gregor,Seville}, attrition~\cite{ZhangGhadiri},
erosion~\cite{Lyczkowski} and fragmentation~\cite{Lecoq}. These phenomena may
occur simultaneously or respectively when an individual particle impacts with
another. After impact separation~\cite{Painter} or
clusterisation~\cite{Clement0} of the two particles occurs. In addition, the
particles may gain or loss mass. Here we will focus on the dynamics of the
collision process which may be decomposed into impact and contact processes.
However, as the contact process is formed, we can also notice
rebound~\cite{Painter} or static contact~\cite{Geng}, or permanent contact,
called cohesion~\cite{Gregor}. These processes exist simultaneously when we
analyse the dynamics of colliding particles. With regard to the granular
dynamics involving many particles in motion, we can observe multiparticle
collisions~\cite{Rioual}, especially when particle concentration is very dense,
because collisional times between several binary particle contacts are higher
in comparison to their separation times. Multiparticle collisions occur when an
individual particle collides with neighbouring particles, so that those
contacts have direct a~influence on each other. Only, an infinitesimally short
collisional time is required for binary collisions~\cite{Painter}. In all the
considered cases the collision process between the two particles is
characterised through the collisional time, which is dependent on the impact
energy and the physical properties of the contacting surfaces. Moreover, after
impact dissipation of energy occurs between the colliding particles. Therefore
the simulations of such dynamics are limited by assumptions concerning the
collision process. One of the major aspects which needs to be taken into account
in the simulations is how to control (correlate) the collisional time and the
energy dissipation associated with an individual particle during the dynamics of
multiparticle collisions.

Generally two different ways exist to model the dynamics of a~granular material.
The continuum approach~\cite{Gidaspow} is based on binary collisions of smooth
spherical particles. Unfortunately, the introduction of real quantities such as
distribution of particle dimensions, particle shapes, their surface wetness and
roughness, etc., greatly limit the application of continuum models. Balzer et
al~\cite{Balzer} inform us that the kinetic theory is useful for the modelling
of gas-solid flow applications in industry: where the geometry involved is
complex (many different inlets or/and outlets). However, the kinetic theory
cannot reflect the real dynamics involved in multiparticle collisions because
the collisional time is defined only for binary collisions.

The discrete deterministic approach more realistically reflects the collision
process. Note that multiparticle collisions in the discrete approach are
decomposed into several binary collisions. In this approach one may distinguish
two general methods. The molecular dynamics method~\cite{Rapaport} takes into
account an expression for the repulsive force acting between a~pair of
contacting particles. In this method particles virtually overlap when a~contact
occurs. The overlap reflects the quantitative deformations of particle surfaces
because the modelling of realistic deformations would be much too complicated.
The interaction laws~\cite{Cundall,Kuwabara,Walton} in the molecular dynamics
method define basic models of the repulsive force for two colliding particles.
They are valid for particle collisions which are independent from one another.
The next method, called the event driven method~\cite{Lubachevsky}, assumes
instantaneous changes in the direction and value of particle velocities
according to conservation equations each time a~binary contact occurs. As shown
in~\cite{Luding0} the basic difference between the event driven and molecular
dynamics methods is the collisional time between a~pair of particles. In the
event driven method this time is ideally zero. Note that this situation is quite
different for the molecular dynamics method, where the contact time is greater
than zero and is dependent on parameters describing the structure of two
contacting surfaces, and is of course dependent on the impact energy. However,
the repulsive force models in the molecular dynamics method underestimate the
energy dissipation in multiparticle collisions~\cite{Luding,Pourin} (This is the
so called ``detachment effect'') but in the event driven method an inelastic
collapse~\cite{McNamara} occurs.

In this paper we will focus on the molecular dynamics method because this gives
us a~chance to correlate the collisional time and the energy dissipation during
multiparticle collisions. We shall introduce a~novel mathematical description of
this method taking into account the division of the collision process into an
impact phase, a~contact phase and another phase occurring after the contact
phase. We assume that the impact phase and the phase formed after the contact
phase are infinitesimally short in time. Consequently, we will analyse the
well-known interaction laws of the repulsive force in the contact phase in
order to examine several difficulties within the collisions. 
On the base of preliminary
results~\cite{Leszczynski} we shall introduce a~novel form of the repulsive
force defined under fractional calculus~\cite{Oldham}. We will also demonstrate
the basic properties of this force and focus on what happens with the
collisional time and the energy dissipation for multiparticle collisions. This
analysis is necessary in computational simulations of the cluster dynamics.
Within the cluster one may notice non-permanent contact and/or cohesion
phenomena between several pairs of colliding particles.

\section{The discrete model of motion of for individual particle}
Let us turn our attention to a~set of spherical particles moving under arbitrary
extortion. The spherical shape of the particle makes only the mathematical
description easier and does not make the model in any way poorer. The reader may
find in~\cite{Matuttis} more information concerning the molecular dynamics
technique adapted to arbitrary form particle shapes. The particles are numbered
by the discrete index $i=1,\ldots,np$, where $np$ is the total number of
considered particles. We describe an individual particle through its radius
$r_{i}$ (or diameter $d_{i}$), mass $m_{i}$, inertia moment ${\cal J}_{i}$,
position $\mathbf{x}_{i}$ of the mass centre, linear speed
$\mathbf{\dot{{x}}}_{i}$ and angular velocity \(\mbox{\boldmath{$\omega$}}_i\).
With regard to the collision of two individual particles we also introduce the
natural function $j(i)$ ($j(i)\neq i$ by assumption) of a~particle $i$ in order
to find the particle index of a~particle in a~set of particles $np$. Several
papers~\cite{Allen,Iwai,Schinner} present different algorithms that detect
particle collisions, being dependent on their shapes, and consequently that to
find the natural function $j(i)$. For a~binary collision we neglect phenomena
which cause a~change in the mass of an individual particle. Thus in our discrete
model we do not take into account fragmentation, attrition and erosion which
eventually take place during the collision process. These phenomena will be the
subject of future investigations.

However, after the contact, which is the second phase of the collision process,
rebound, non-permanent contact (static contact) or cohesion can arise
simultaneously. In this paper, we will try to model above the phenomena by
introducing a~novel mathematical description and a~novel form of the repulsive
force into the molecular dynamics method.

\subsection{Mapping local coordinates onto global ones and defining the overlap}
\begin{figure}[ht] 
 \begin{center}
  \includegraphics[width=0.80\columnwidth,keepaspectratio]{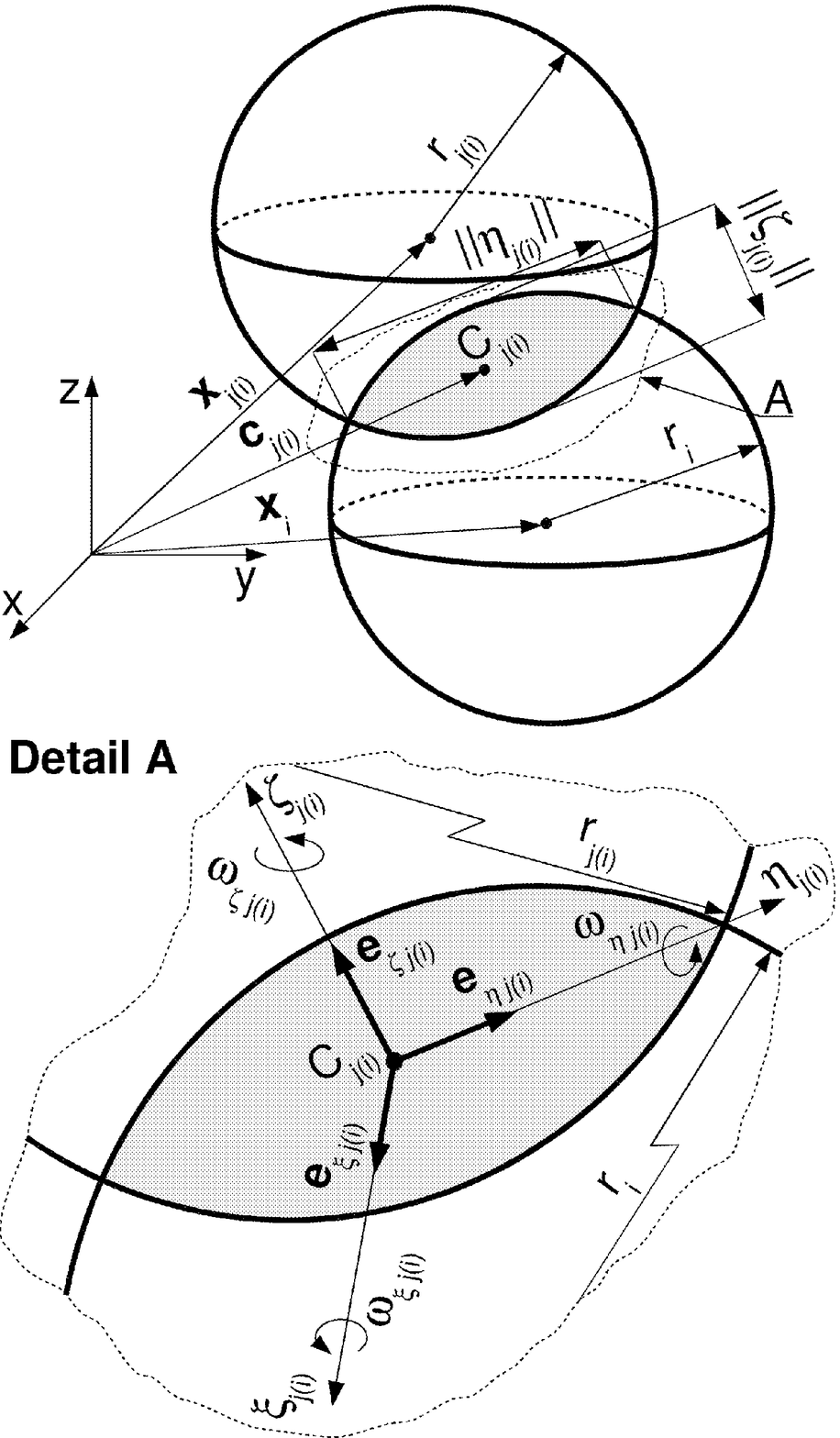}
 \end{center}
 \caption{Scheme illustrates particle collisions with useful notations.
          \label{fig01}}
\end{figure}

Starting from the description shown in Fig.~\ref{fig01}, let us introduce
several definitions before formulating the motion equations.
First, we assign local coordinates as $(\xi,\eta,\zeta)$ and global ones as
$(x,y,z)$. When we consider a~contact which eventually takes place between
two particles then the normal unit vector $\mathbf{e}_{\zeta\, j(i)}$ that
connects the particle's centres of mass reads
\begin{equation} \label{eq01}
 \mathbf{e}_{\zeta\, j(i)}=\frac{\mathbf{x}_{j(i)}-\mathbf{x}_{i}}{\left\Vert
  \mathbf{x}_{j(i)}-\mathbf{x}_{i}\right\Vert }=
  \left[e_{\zeta\, j(i)}^{x},e_{\zeta\, j(i)}^{y},e_{\zeta\, j(i)}^{z}\right],
\end{equation}
where $\Vert\cdot\Vert$ represents a~norm calculated from the relative
coordinate $\mathbf{x}_{j(i)}-\mathbf{x}_{i}$. Tangential unit vectors which
operate on a~tangent plane (rotated by $\frac{\pi}{2}$ to the normal) become
\begin{equation} \label{eq02}
 \mathbf{e}_{\eta\, j(i)}=
 \left[e_{\zeta\, j(i)}^{y},-e_{\zeta\, j(i)}^{x},0\right]
 \frac{\left\Vert \mathbf{x}_{j(i)}-\mathbf{x}_{i}\right\Vert }
      {\left\Vert \mathbf{x}_{j(i)}-\mathbf{x}_{i}\right\Vert _{x,y}},
\end{equation}
\begin{equation} \label{eq03}
 \mathbf{e}_{\xi\, j(i)}=\mathbf{e}_{\eta\, j(i)}\times
 \mathbf{e}_{\zeta\, j(i)},
\end{equation}
where $\Vert\cdot\Vert_{x,y}$ represents the norm which is calculated only in
the tangent plane. When a~particle hits a~wall we redefine unit
vectors~(\ref{eq01}), (\ref{eq02}) and (\ref{eq03}) putting
$\mathbf{xb}_{n(i)}$ instead of $\mathbf{x}_{j(i)}$, where $\mathbf{xb}_{n(i)}$
is a~point whose coordinates issue from the line that crosses the particle's
centre of mass and is perpendicular to the wall. The general form of the base
vectors is presented as follows
\begin{equation} \label{eq003}
 \mathbf{e}_{j(i)}=\left[\begin{array}{l}
 \mathbf{e}_{\xi\, j(i)}\\
 \mathbf{e}_{\eta\, j(i)}\\
 \mathbf{e}_{\zeta\, j(i)}\end{array}\right].
\end{equation}

Fig.~\ref{fig01} presents the ``virtual overlap''
$\left\Vert \mbox{\boldmath{$\zeta$}}_{j(i)}\right\Vert$
of two particles experiencing a~contact. With regard to~\cite{Herrmann}
we define the overlap as
\begin{equation} \label{eq04}
 \left\Vert \mbox{\boldmath{$\zeta$}}_{j(i)}\right\Vert = 
 r_{j(i)} + r_i - \left\Vert \mathbf{x}_{j(i)} - \mathbf{x}_i \right\Vert,
\end{equation}
which is associated with the particles having spherical forms. Note that only
positive values of formula~(\ref{eq04}) indicate a~contact while negative ones
confirm that the considered particles are in separation. This means that they
move individually. As presented in previous section and in Fig.~\ref{fig01}, the
overlap reflects the penetration depth of the particles in a~direction which
connects the particle's centres of mass, pointing from $i$ to $j(i)$. We also
introduce the penetration width of the particles defined as the direction
perpendicular to the previous one. Thus we have
\begin{eqnarray} \label{eq05}
 && \left\Vert \mbox{\boldmath{$\eta$}}_{j(i)}\right\Vert = 
 \left\Vert \mbox{\boldmath{$\xi$}}_{j(i)}\right\Vert \nonumber \\ 
 &=& 2\sqrt{r_i^2 - \left(r_i + \frac{1}{2}\left\Vert 
 \mbox{\boldmath{$\zeta$}}_{j(i)}\right\Vert
 \frac{2 r_{j(i)} - \left\Vert \mbox{\boldmath{$\zeta$}}_{j(i)}\right\Vert}
      {\left\Vert \mbox{\boldmath{$\zeta$}}_{j(i)}\right\Vert-
      \left(r_i + r_{j(i)}\right)}
 \right)^2}
\end{eqnarray}
valid for \(\left\Vert \mbox{\boldmath{$\zeta$}}_{j(i)}\right\Vert\geq 0 \).
Let $\mathbf{c}_{j(i)}$ be a~vector which defines a~point $C_{j(i)}$ of the
application of the repulsive force, and which is taken as the mass centre of the
overlapping region~(\ref{eq04}) as shown in Fig.~\ref{fig01}. Taking into
consideration the fact the that particles have only spherical forms and collide
when their overlap~(\ref{eq04}) is positive we obtain
\begin{equation} \label{eq06}
 \mathbf{c}_{j(i)} = \mathbf{x}_i +
 \left( r_i - \frac{\left\Vert \mbox{\boldmath{$\zeta$}}_{j(i)}\right\Vert 
 \left(r_{j(i)} - \left\Vert \mbox{\boldmath{$\zeta$}}_{j(i)}\right\Vert\right)}
 {r_i + r_{j(i)} -\left\Vert \mbox{\boldmath{$\zeta$}}_{j(i)}\right\Vert}\right)
 \mathbf{e}_{\zeta\,j(i)}.
\end{equation}
Thus, at the beginning of a~collision we have
\begin{equation} \label{eq07}
 \mathbf{c}_{j(i)} = \frac{r_i \mathbf{x}_{j(i)} + r_{j(i)} 
 \mathbf{x}_i}{r_{j(i)}+r_i},
\end{equation}
and one can find an explicit time $t_{j(i)}^{b}$ where the overlap~(\ref{eq04})
is zero. Above notation allows us to analyse multiparticle collisions where an
individual particle $i$ collides with neighbouring particles $j(i)$. Therefore
many overlaps~(\ref{eq04}) indexed $j(i)$ on the particle $i$ may occur. This
allows us to formulate the motion equations in the right form. 

Next we introduce the relative velocity of a~particle $i$ and a~particle $j(i)$
at point $C_{j(i)}$ as
\begin{eqnarray} \label{eq10}
 \mathbf{\widetilde{u}}_{j(i)} &=& \mathbf{\widetilde{u}}_{j(i)}^{lin} +
 \mathbf{\widetilde{u}}_{j(i)}^{rot} =  \mathbf{\dot{x}}_i -
 \mathbf{\dot{x}}_{j(i)} \nonumber \\
 &-& \left( \mbox{\boldmath{$\omega$}}_i \times \mathbf{s}_{i,j(i)} +
 \mbox{\boldmath{$\omega$}}_{j(i)} \times \mathbf{s}_{j(i),i}\right),
\end{eqnarray}
where $ \mathbf{\widetilde{u}}_{j(i)}^{lin}$ and
$ \mathbf{\widetilde{u}}_{j(i)}^{rot} $ are linear and rotational relative
velocities and $\mathbf{s}_{i,j(i)}$, $\mathbf{s}_{j(i),i}$ are branch vectors
connecting the mass centres of particles $i$ and $j(i)$ with the point
$C_{j(i)}$ of application of the repulsive force. Note that above values are
defined for the global system of coordinates $(x,y,z)$. To change this to the
local system of coordinates we need to use the scalar product of the base
vectors~(\ref{eq003}) per the vector of the relative velocity. Calculating
the branch vectors we obtain the following dependencies in the local system
$(\xi,\eta,\zeta)$ as
\begin{equation} \label{eq11}
 \mathbf{s}_{i,j(i)}^{'} =\left[ 0,0,-\left\Vert 
 \mbox{\boldmath{$\widetilde{\zeta}$}}_{j(i)} \right\Vert\right], \,
 \mathbf{s}_{j(i),i}^{'} =\left[ 0,0, \left\Vert
 \mbox{\boldmath{$\widetilde{\widetilde{\zeta}}$}}_{j(i)} \right\Vert\right],
\end{equation}
where
\begin{eqnarray} \label{eq12}
 \left\Vert \mbox{\boldmath{$\widetilde{\zeta}$}}_{j(i)} \right\Vert &=&
 \left\Vert \mathbf{c}_{j(i)} - \mathbf{x}_i \right\Vert \nonumber \\
 &=& r_i - \left\Vert \mbox{\boldmath{$\zeta$}}_{j(i)} \right\Vert
 \frac{\frac{1}{2}\left\Vert \mbox{\boldmath{$\zeta$}}_{j(i)}
 \right\Vert - r_{j(i)}}{\left\Vert \mbox{\boldmath{$\zeta$}}_{j(i)} \right\Vert
 - \left( r_i + r_{j(i)}\right)},
\end{eqnarray}
\begin{eqnarray} \label{eq13}
 \left\Vert \mbox{\boldmath{$\widetilde{\widetilde{\zeta}}$}}_{j(i)} \right\Vert
 &=& \left\Vert \mathbf{c}_{j(i)} - \mathbf{x}_{j(i)} \right\Vert \nonumber \\
 &=& r_{j(i)} - \left\Vert \mbox{\boldmath{$\zeta$}}_{j(i)} \right\Vert
 \frac{\frac{1}{2} \left\Vert \mbox{\boldmath{$\zeta$}}_{j(i)} \right\Vert - r_i}
      {\left\Vert \mbox{\boldmath{$\zeta$}}_{j(i)} \right\Vert -
      \left( r_i + r_{j(i)}\right)}.
\end{eqnarray}
In the global system of coordinates the branch vectors become
\begin{equation} \label{eq013}
 \mathbf{s}_{i,j(i)}=\mathbf{e}\cdot\mathbf{s}_{i,j(i)}^{'\, T},\qquad
 \mathbf{s}_{j(i),i}=\mathbf{e}\cdot\mathbf{s}_{j(i),i}^{'\, T}.
\end{equation}
Using Eqs.~(\ref{eq003}) and~(\ref{eq10}) we translate the linear and rotational
relative velocities in the global system of coordinates to the local one as
\begin{equation} \label{eq0013}
 \widetilde{\mathbf{u}}_{j(i)}^{'\, lin}=\mathbf{e}\cdot
 \widetilde{\mathbf{u}}_{j(i)}^{lin},\quad
 \widetilde{\mathbf{u}}_{j(i)}^{'\, rot}=
 \mathbf{e}\cdot\widetilde{\mathbf{u}}_{j(i)}^{rot},
\end{equation}
where $\widetilde{u}_{\zeta\, j(i)}^{lin}$, $\widetilde{u}_{\zeta\, j(i)}^{rot}$
are the relative velocities operating in the normal direction to the contacting
surfaces as shown on Fig.~\ref{fig01} and
$
 \widetilde{\mathbf{u}}_{t\, j(i)}^{'\, lin}=\left[
 \widetilde{u}_{\xi\, j(i)}^{lin},\,\widetilde{u}_{\eta\, j(i)}^{lin}\right]
$,
$
 \widetilde{\mathbf{u}}_{t\, j(i)}^{'\, rot}=\left[
 \widetilde{u}_{\xi\, j(i)}^{rot},\,\widetilde{u}_{\eta\, j(i)}^{rot}\right]
$
denote vectors of the relative velocities acting in the tangential direction
(rotated by $\frac{\pi}{2}$ to the normal). Additionally we use the same
translation as presented by expression~(\ref{eq0013}) for calculations
$
 \mbox{\boldmath{$\omega$}}_i^{'} = \mathbf{e}\cdot\mbox{\boldmath{$\omega$}}_i,
$
$
 \mbox{\boldmath{$\omega$}}_{j(i)}^{'} = \mathbf{e}\cdot
 \mbox{\boldmath{$\omega$}}_{j(i)}
$
in order to obtain the angular velocities for particles $i$ and $j(i)$ in the
local system of coordinates $(\xi,\eta,\zeta)$.

If a~collision between a~particle and a~wall takes place, the
overlap~(\ref{eq04}) is defined as
\begin{equation} \label{eq16}
 \left\Vert \mbox{\boldmath{$\zeta$}}_{j(i)}^{\mbox{b}}\right\Vert = 
 r_i - \left\Vert \mathbf{xb}_{j(i)} - \mathbf{x}_i \right\Vert,
\end{equation}
and also we have
\begin{equation} \label{eq17}
 \left\Vert \mbox{\boldmath{$\eta$}}_{j(i)}^{\mbox{b}}\right\Vert = 2 
 \sqrt{\left\Vert \mbox{\boldmath{$\zeta$}}_{j(i)}^{\mbox{b}}\right\Vert
 \left( 2 r_i - \left\Vert \mbox{\boldmath{$\zeta$}}_{j(i)}^{\mbox{b}}
 \right\Vert\right)}
\end{equation}
which is valid for
$\left\Vert \mbox{\boldmath{$\zeta$}}_{j(i)}^{\mbox{b}}\right\Vert\geq 0$.
In this case the point of application $\mathbf{c}_{j(i)}^{b}$ is defined by the
following formula
\begin{equation} \label{eq18}
 \mathbf{c}_{j(i)}^{\mbox{b}} = \mathbf{x}_i + \left( r_i -\frac{5}{8}\left\Vert
 \mbox{\boldmath{$\zeta$}}_{j(i)}^{\mbox{b}}\right\Vert\right)
 \mathbf{e}_{j(i)}^{\mbox{b}},
\end{equation}
where
\begin{equation} \label{eq19}
 \mathbf{e}_{j(i)}^{\mbox{b}} = \frac{\mathbf{xb}_{j(i)} - \mathbf{x}_i }
 {\left\Vert \mathbf{xb}_{j(i)} - \mathbf{x}_i \right\Vert}
\end{equation}
becomes a~normal unit vector which is perpendicular to the wall. When
a~particle-wall collision begins we obtain
$\mathbf{c}_{j(i)}^{b}=\mathbf{x}_{i}+r_{i}\mathbf{e}_{\zeta\, j(i)}^{b}$.
Moreover, one can find the explicit time $t_{j(i)}^{bb}$ when the
overlap~(\ref{eq16}) is zero. Expressions~(\ref{eq10})$-$(\ref{eq0013}), defined
for a~particle-particle collision, may be redefined in simple way for
a~particle-wall
collision when $\mathbf{\dot{x}}_{j(i)}$ and $\mbox{\boldmath{$\omega$}}_{j(i)}$
are zeros and unit vectors are also redefined as explained in previous
considerations. For example a~component of the branch vector~(\ref{eq12}) is
redefined for a~particle-wall collision and takes the following form
\begin{equation} \label{eq20}
 \left\Vert \mbox{\boldmath{$\widetilde{\zeta}$}}_{j(i)}^{\mbox{b}} \right\Vert=
 r_i - \frac{5}{8} \left\Vert \mbox{\boldmath{$\zeta$}}_{j(i)}^{\mbox{b}}
 \right\Vert.
\end{equation}
We neglect here any additional expressions necessary to describe the
particle-wall collision. The reader can do this very easily in the same way
as explained previously.

Summarising our considerations, we introduced the above mathematical description
which is necessary for the formulation of the motion equations and is also
necessary for some forms of the repulsive force, acting for both
particle-particle and particle-wall collisions.

\subsection{Motion equations}
The molecular dynamics method requires a~discrete deterministic approach in
order to the model motion of an individual particle. Note that the particle may
collide or
lose contact with other particles. Therefore in motion equations we need to add
or reject some forms of the repulsive force and/or the attractive force in order
to simulate the particle dynamics more realistically. In this paper we neglect
the attractive force and we will concentrate only on the repulsive force. 
Aganist this background, let us describe the motion of an individual particle
by the following two sets of equations
\begin{equation} \label{eq21}
 \left\{
 \begin{array}{c}  m_i \mathbf{\ddot{x}}_i = \sum\limits_l \mathbf{F}_l\\
 {\cal J}_i \mbox{\boldmath{$\dot{\omega}$}}_i = \sum\limits_l \mathbf{M}_l
 \end{array}
 \right.
\end{equation}
suitable for particle motion without any collision, and
\begin{equation} \label{eq22}
 \left\{\begin{array}{lcl}  m_i \mathbf{\ddot{x}}_i & = &
 \sum\limits_{j(i), j(i)\neq i} \mathbf{P}_{j(i)}^{coll} \\ & + &
 \sum\limits_{j(i), j(i)\neq i} \mathbf{P}_{j(i)}^{b \; coll} + 
 \sum\limits_l \mathbf{F}_l \\ {\cal J}_i \mbox{\boldmath{$\dot{\omega}$}}_i
 & = & \sum\limits_{j(i), j(i)\neq i} \mathbf{M}_{i,j(i)}^{coll} \\
 & + & \sum\limits_{j(i), j(i)\neq i} \mathbf{M}_{i,j(i)}^{b \; coll} +
 \sum\limits_l \mathbf{M}_l \end{array} \right.,
\end{equation}
which takes into account multiparticle collisions. The above sets of equations
exist simultaneously over time and are dependent on the detection of a~contact
and the administration of the repulsive force-overlap path during the contact.
In both
Eqs.~(\ref{eq21}) and~(\ref{eq22}) $\mathbf{F}_{l}$ denotes an arbitrary force
which extorts the motion of a~particle, $\mathbf{M}_{l}$ is an arbitrary torque,
$\mathbf{P}_{j(i)}^{coll}$ is a~collisional force composed of the repulsive
and friction forces and acts between a~pair of colliding particles,
$\mathbf{P}_{j(i)}^{b\; coll}$ is also the collisional force operating on
a~particle-wall collision, $\mathbf{M}_{i,j(i)}^{coll}$ and
$\mathbf{M}_{i,j(i)}^{b\; coll}$ are collisional torques definitively operating
on particle-particle and particle-wall collision.

We need to define some of the criteria necessary for handling the above two sets
of equations over the time of the calculations. It was evidently shown in the
previous
subsection that for the beginning of a~collision the overlap given by
expression~(\ref{eq04}) or~(\ref{eq16}) is zero. Thus we have the impact phase.
However, some of the criteria for determining when the collision ends are
unclear. Correctly predicting the separation time of two colliding particles is
crucial in the calculation. Most papers assume the particles separate at the
time when the
overlap returns to zero. As proved in~\cite{Zhang0}, the repulsive force changes
direction at the time when the overlap returns to zero. This is contrary to
experimental evidence, also shown and compared with some models
in~\cite{Zhang0}, when the force does not change direction. An
attractive force operating in opposite direction to the repulsive force has
different origins and is not taken into account here.

At this crucial point of our considerations, we need to introduce some
definitions in
order to predict correctly the beginning time of a~particle collision and the
time when the collision ends. Let us consider the time of calculations
$t\in\left\langle 0,T\right\rangle $ where $T$ represents the total time in
which the calculations are performed. We also define the time step $\Delta t$ in
which we trace the system dynamics. Follow on from previous explanations we
start with some conditions.
\begin{definition}
 If, within a~time interval $\left\langle t,t+\Delta t\right\rangle$ detects the
 beginning of a~collision between a~pair of particles is detected then the
 overlap~(\ref{eq04}) should fulfil the following conditions
 \begin{equation} \label{eq23}
  \begin{array}{c}
   \left\Vert \mbox{\boldmath{$\zeta$}}_{j(i)}(t)\right\Vert \leq 0 \hspace{3mm}
   \mbox{and} \hspace{3mm} \left\Vert 
   \mbox{\boldmath{$\zeta$}}_{j(i)}(t + \Delta t)\right\Vert \geq 0 \\
   \mbox{and therefore} \hspace{3mm} \left\Vert \mbox{\boldmath{$\zeta$}}_{j(i)}
   \left( t_{j(i)}^{b} \right)\right\Vert = 0,
  \end{array}
\end{equation}
and then time $t_{j(i)}^{b}\in\left\langle t,t+\Delta t\right\rangle$ is the
time when the collision starts.
\end{definition}
\begin{definition}
 If, within \ a~time \ interval \ $\left\langle t,t+\Delta t\right\rangle $
 the end of a~collision is formed then the overlap~(\ref{eq04}) and the normal
 component of the repulsive force $R_{\zeta\, j(i)}$ should fulfil the following
 conditions
 \begin{equation} \label{eq24}
  \begin{array}{c}
   \left\Vert \mbox{\boldmath{$\zeta$}}_{j(i)}(t)\right\Vert \geq 0 \hspace{3mm}
   \mbox{and} \hspace{3mm} \left\Vert
   \mbox{\boldmath{$\zeta$}}_{j(i)}(t + \Delta t)\right\Vert \leq 0 \\
   \mbox{and} \hspace{3mm} R_{\zeta\, j(i)}(t) > 0 \hspace{3mm} \mbox{and}
   \hspace{3mm} R_{\zeta\, j(i)}(t + \Delta t) > 0 \\ \mbox{and therefore}
   \hspace{3mm} \left\Vert
   \mbox{\boldmath{$\zeta$}}_{j(i)}\left( t_{j(i)}^{e} \right)\right\Vert = 0
  \end{array}
 \end{equation}
 or
 \begin{equation} \label{eq25}
  \begin{array}{c}
   \left\Vert \mbox{\boldmath{$\zeta$}}_{j(i)}(t)\right\Vert > 0 \hspace{3mm}
   \mbox{and} \hspace{3mm} \left\Vert
   \mbox{\boldmath{$\zeta$}}_{j(i)}(t + \Delta t)\right\Vert > 0 \\ \mbox{and}
   \hspace{3mm} R_{\zeta\, j(i)}(t) \geq 0 \hspace{3mm} \mbox{and} \hspace{3mm}
   R_{\zeta\, j(i)}(t + \Delta t) \leq 0 \\ \mbox{and therefore} \hspace{3mm}
   R_{\zeta\, j(i)}\left(t_{j(i)}^{e} \right) = 0,
  \end{array}
 \end{equation}
 and then time $t_{j(i)}^{e}\in\left\langle t,t+\Delta t\right\rangle $ is the
 time when the collision ends.
\end{definition}
In formulae~(\ref{eq24}) and~(\ref{eq25}) $R_{\zeta\, j(i)}$ represents a~normal
component of the repulsive force. In the next subsection we will introduce
a~definition of this force.
\begin{definition}
 If, within a~time interval $\left\langle t,t+\Delta t\right\rangle $ the
 overlap and the normal component of the repulsive force $R_{\zeta\, j(i)}$
 behave as follows
 \begin{equation} \label{eq26}
  \begin{array}{c}
   \left\Vert \mbox{\boldmath{$\zeta$}}_{j(i)}(t)\right\Vert > 0 \hspace{3mm}
   \mbox{and} \hspace{3mm} \left\Vert
   \mbox{\boldmath{$\zeta$}}_{j(i)}(t + \Delta t)\right\Vert > 0 \\ \mbox{and}
   \hspace{3mm} R_{\zeta\, j(i)}(t + \Delta t) \rightarrow 0^{+},
  \end{array}
 \end{equation}
 then time $t_{j(i)}^{e}=\Delta t+t$ is the time when the collision ends.
\end{definition}
\begin{definition}
 When the condition~(\ref{eq24}) is fulfilled then linear and rotational
 components~(\ref{eq0013}) of the relative velocity predict the following states
 after the collision:
 \begin{itemize}
  \item rebound of particles without particle deformations for
        $\widetilde{u}_{\zeta\, j(i)}^{lin}\left(t_{j(i)}^{e}\right)\neq0$
        and $\widetilde{u}_{\zeta\, j(i)}^{lin}\left(t_{j(i)}^{e}\right)$
        has an opposite direction (sign) to
	$\widetilde{u}_{\zeta\, j(i)}^{lin}\left(t_{j(i)}^{b}\right)$,
  \item torsion for
        $
	 \omega_{\zeta\, i}\left(t_{j(i)}^{e}\right)-
	 \omega_{\zeta\, j(i)}\left(t_{j(i)}^{e}\right)\neq0
	$
        or sliding for
	$
	 \widetilde{\mathbf{u}}_{t\, j(i)}^{'\, lin}\left(t_{j(i)}^{e}\right)
	 \neq\mathbf{0}
	$
        or rolling for
	$
	 \widetilde{\mathbf{u}}_{t\, j(i)}^{'\, rot}\left(t_{j(i)}^{e}\right)
	 \neq\mathbf{0}
	$
        of particles without particle deformations for
	$\widetilde{u}_{\zeta\, j(i)}^{lin}\left(t_{j(i)}^{e}\right)=0$,
  \item non-permanent \ or permanent stick \ of particles without \ particle
        deformations \ for \ 
	$
	 \widetilde{u}_{\zeta\, j(i)}^{lin}\left(t_{j(i)}^{e}\right)\, + \,
	 \widetilde{u}_{\zeta\, j(i)}^{rot}\left(t_{j(i)}^{e}\right)\, =\, 0
	$
        \hspace{3mm} and for \hspace{3mm} 
	$
	 \widetilde{\mathbf{u}}_{t\, j(i)}^{'\, lin}\left(t_{j(i)}^{e}\right)\,
	 + \,
	 \widetilde{\mathbf{u}}_{t\, j(i)}^{'\, rot}\left(t_{j(i)}^{e}\right)=
	 \mathbf{0}
	$. 
 \end{itemize}
\end{definition}
\begin{definition}
 When the condition~(\ref{eq25}) is fulfilled then components~(\ref{eq0013}) of
 the relative velocity predict the same states as described by definition~4 but
 particle deformations are noted.
\end{definition}
\begin{definition}
 When the condition~(\ref{eq26}) is fulfilled then the normal component
 $\widetilde{u}_{\zeta\, j(i)}^{lin}\left(t_{j(i)}^{e}\right)$ of the relative
 velocity~(\ref{eq0013}) predicts adhesion-induced plastic deformations of
 particles or breakage of particles depending on the hardness of the contacting
 surfaces. 
\end{definition}

On the base of previous assumptions and definitions~1 and 2
we
introduce the collisional time between a~pair of contacting particles as
$t_{j(i)}^{coll}=t_{j(i)}^{e}-t_{j(i)}^{b}$. This time is determined by
conditions~(\ref{eq24}) and~(\ref{eq25}) simultaneously. In other words, when
the overlap changes sign faster than the repulsive force changes direction or
vice versa then the collision is finished. If particles are still in contact then
the total contact time is significantly greater than the collisional time. If
particles are separated then the total contact time equals the collisional
time. As presented in the first section, the collisional process composes
the impact phase, the contact phase and the last phase formed after the contact
phase. Moreover, when the formulation of the first and the last phases is
infinitesimally short in time then the collisional time is predicted by the
contact phase. The contact phase is predicted by the repulsive force-overlap
path. The adhesion or cohesion states extend 
the contact phase over time to infinity. In our approach adhesion and cohesion
are eventually formed after the impact and they represent completely different
phenomena which definitely result from the collision process. Generally, when
we model impact dynamics we need to consider the balance between the repulsive
force which is a~direct reaction to the impact, and the attractive forces which
are a~result i.e. cohesion of particles. Therefore, our collisional time
$t_{j(i)}^{coll}$ also becomes the time of relaxation in which the collision
process is stopped and novel states are formed. Most papers neglect this fact
and identify the total contact time, which may increase to infinity, as the
collisional one.

Extending our considerations we notice that definition~4 is suitable for the
elastic
collisions of particles because there are no deformations in contacting
particles
- the overlap tends to zero faster than the repulsive force changes direction.
In definition~5 we observe the opposite situation - the repulsive force changes
direction faster than the overlap tends to zero.
In our approach the
collision process is fully controlled by the repulsive force except for the
situation
presented by formula~(\ref{eq26}) in definition~3. On the basis of definition~6,
which results from definition~3, we are able to explain that the local stresses
associated with deformations of contacting particles become sufficiently large
so as to exceed the elastic limit of the materials to a~result plastic flow
occurs~\cite{Krupp} and the behaviour of particle adhesion differs from that
predicted by the elastic deformation theory~\cite{Maugis}. Adhesion-induced plastic
deformations of contacting materials are evidently shown in some
experiments~\cite{Rimai}. Therefore we have two possible states resulting from
the impact: particle clusterisations when the colliding materials are soft, and
fragmentation of particles when the colliding materials are hard.

Summarising this subsection we formulated two general forms of the motion
equations and discussed precisely how to handle them.

\subsection{Collisional forces, collisional torques and the fractional
            interaction law}
With regard to motion equations~(\ref{eq22}) we introduce a~mathematical
description of the collisional forces and torques occurring in such a~system.
In the normal direction to the contacting surfaces we apply only a~repulsive
force,
completely neglecting any attractive forces. In~\cite{Seville} one can find some
forms of attractive forces and their physical meanings. In a~tangential plane we
introduce a~system of friction forces and torques. According to the friction
mechanism, the tangential friction force is one of four types: torsion with
sliding friction, sliding friction, rolling friction or static friction.
Torsion friction occurs when colliding particles differ by their angular
velocities in the normal direction $\omega_{\zeta\, i}$ and
$\omega_{\zeta\, j(i)}$. Torsion with sliding friction is for colliding
particles which have different angular velocities in the normal direction and
different linear velocities in the tangential plane. Sliding friction happens
when
slipping occurs in colliding particles. When the relative linear velocity of
the particles in the tangent direction reduces to zero, sliding friction is
replaced by rolling friction. If the external forces are sufficiently small,
the rolling friction reduces the velocity until particle motion stops and static
friction occurs. Considering the impact dynamics, we implemented the following
mechanism in general form: torsion with sliding friction can change to rolling
friction and the rolling friction tends to static friction. More details
concerning the modelling of torsion, sliding and rolling friction can be found
in~\cite{Farkas,Zhang1}. Here we show a~description of the collisional force in
global coordinates $(x,y,z)$ as
\begin{equation} \label{eq28}
 \mathbf{P}_{j(i)}^{coll}=\left\{
  \begin{array}{lll}
   \mathbf{P}_{j(i)}^{sta} & \mbox{for} & 
    \left\Vert \widetilde{\mathbf{u}}_{t\, j(i)}^{'\, lin}\right\Vert =
    \left\Vert \widetilde{\mathbf{u}}_{t\, j(i)}^{'\, rot}\right\Vert =0\\
   \mathbf{P}_{j(i)}^{rol} & \mbox{for} &
    \left\Vert \widetilde{\mathbf{u}}_{t\, j(i)}^{'\, lin}-
    \widetilde{\mathbf{u}}_{t\, j(i)}^{'\, rot}\right\Vert =0\\
   \mathbf{P}_{j(i)}^{sli} & \mbox{for} &
    \left\Vert \widetilde{\mathbf{u}}_{t\, j(i)}^{'\, lin}-
    \widetilde{\mathbf{u}}_{t\, j(i)}^{'\, rot}\right\Vert >0
  \end{array}
  \right.,
\end{equation}
where $\mathbf{P}_{j(i)}^{sta}$ is the force acting in a~static friction state,
$\mathbf{P}_{j(i)}^{rol}$ is the force occurring in a~rolling state and
$\mathbf{P}_{j(i)}^{sli}$ is the force coupling the torsion-sliding state. The
emphasis in this paper is on the impact dynamics the static friction is
only implemented in a~simple form. A~more detailed model of the static friction
state requires analysis of the tangential displacement and possibly the
inclusion of time dependent effects.
According to Fig.~\ref{fig01} we need to define the collisional force in the
local system of coordinates $(\xi,\eta,\zeta)$. Using a~matrix of the base
vectors~(\ref{eq003}) we introduce transition from the local system to the
global ones as
\begin{equation} \label{eq29}
 \mathbf{P}_{j(i)}^{sli}=\mathbf{e}_{j(i)}^{T}\cdot\mathbf{P}_{j(i)}^{'\, sli},
 \qquad\mathbf{P}_{j(i)}^{rol}=\mathbf{e}_{j(i)}^{T}\cdot
 \mathbf{P}_{j(i)}^{'\, rol},
\end{equation}
where $\mathbf{P}_{j(i)}^{'\, sli}$, $\mathbf{P}_{j(i)}^{'\, rol}$ are forces
defined in the local system of coordinates as
\begin{equation} \label{eq30}
 \mathbf{P}_{j(i)}^{'\, sli}=\left[
 \begin{array}{l}
  T_{\xi\, j(i)}^{sli}\\ T_{\eta\, j(i)}^{sli}\\ -R_{\zeta\, j(i)}
 \end{array}
 \right],\qquad \mathbf{P}_{j(i)}^{'\, rol}=\left[
 \begin{array}{l}
  T_{\xi\, j(i)}^{rol}\\ T_{\eta\, j(i)}^{rol}\\ -R_{\zeta\, j(i)}
 \end{array}
 \right].
\end{equation}
In expression~(\ref{eq30}) $T_{\xi\, j(i)}^{sli}$, $T_{\eta\, j(i)}^{sli}$,
$T_{\xi\, j(i)}^{rol}$, $T_{\eta\, j(i)}^{rol}$ represent components of the
friction force in a~plane $(\xi,\eta)$ for torsion-sliding and rolling states,
$R_{\zeta\, j(i)}$ is a~sum of the normal components of attractive and repulsive
forces operating during a~collision. As assumed in this paper, we neglect
attractive forces and concentrate only on forms of the repulsive force. Some
forms of the attractive forces can be found in~\cite{Gregor,Seville} but
the most well-known forms of the repulsive force are
in~\cite{Cundall,Kuwabara,Walton}.

On the basis of preliminary results~\cite{Leszczynski} we now introduce a~model
of the repulsive force in general form called the fractional interaction law.
Thus we have
\begin{widetext}
 \begin{equation} \label{eq31}
  R_{\zeta\,j(i)} = \left\{ 
  \begin{array}{lll} 
   \max\left[0, c_{j(i)}^{\alpha_{j(i)}} k_{j(i)}^{1-\alpha_{j(i)}} 
   \,_{t_{j(i)}^b}\!\!{\cal D}_{t_{j(i)}^e}^{\alpha_{j(i)}} \left( \left\Vert 
   \mbox{\boldmath{$\zeta$}}_{j(i)}\right\Vert \right)\right] & \mbox{for} & 
   \left\Vert \mbox{\boldmath{$\zeta$}}_{j(i)}\right\Vert\geq 0 \\ 0 &
   \mbox{for} & \left\Vert \mbox{\boldmath{$\zeta$}}_{j(i)}\right\Vert < 0
  \end{array}
  \right.,
 \end{equation}
\end{widetext}
where $c_{j(i)}$, $k_{j(i)}$ are damping and spring coefficients with the same
meaning as in the linear interaction law~\cite{Cundall},
$\left\Vert \mbox{\boldmath{$\zeta$}}_{j(i)}\right\Vert$ represents the overlap
defined by formula~(\ref{eq04}), $t_{j(i)}^{b}$, $t_{j(i)}^{e}$ are start and
stop times of a~collision (not a~total contact) predicted by several definitions
in the previous subsection, as explained in~\cite{Leszczynski} $\alpha_{j(i)}$
is
the conversion degree of impact energy into viscoelasticity of the material and
$
 _{t_{j(i)}^b}\!\!{\cal D}_{t_{j(i)}^e}^{\alpha_{j(i)}}\left( \left\Vert
 \mbox{\boldmath{$\zeta$}}_{j(i)}\right\Vert\right)
$
represents general form of the differential and integral operator of fractional
order. According to fractional calculus~\cite{Oldham,Samko} we introduce the
definition of this operator in the following form
\begin{widetext}
 \begin{equation} \label{eq32}
 _{t_{j(i)}^b}\!\!{\cal D}_{t_{j(i)}^e}^{\alpha_{j(i)}}\left( \left\Vert
 \mbox{\boldmath{$\zeta$}}_{j(i)}(t)\right\Vert\right) = \left\{
 \begin{array}{lcl}
  \sum\limits_{l=0}^{n-1}\frac{\left(t-t_{j(i)}^b\right)^{l-\alpha_{j(i)}}}
  {\Gamma\left(l-\alpha_{j(i)}+1\right)} \left\Vert
  \mbox{\boldmath{$\zeta$}}_{j(i)}^{(l)}\left(t_{j(i)}^b\right)\right\Vert 
  +\;_{t_{j(i)}^b}^{C}\!\!D_{t_{j(i)}^e}^{\alpha_{j(i)}}\left( \left\Vert
  \mbox{\boldmath{$\zeta$}}_{j(i)}(t)\right\Vert\right) & \mbox{for} &
  \alpha_{j(i)}\geq 0 \\ _{t_{j(i)}^b}\!\!I_{t_{j(i)}^e}^{-\alpha_{j(i)}}\left(
  \left\Vert \mbox{\boldmath{$\zeta$}}_{j(i)}(t)\right\Vert\right) & \mbox{for}
  & \alpha_{j(i)}<0 \end{array}\right.,
 \end{equation}
\end{widetext}
where \ $t$ \ denotes actual time of calculations
$t\in\left\langle t_{j(i)}^{b},t_{j(i)}^{e}\right\rangle $, the sum represents
the initial conditions,
$
 _{t_{j(i)}^b}^{C}\!\!D_{t_{j(i)}^e}^{\alpha_{j(i)}}\left( \left\Vert
 \mbox{\boldmath{$\zeta$}}_{j(i)}(t)\right\Vert\right)
$
is the Caputo fractional derivative
\begin{widetext}
 \begin{equation} \label{eq33}
  _{t_{j(i)}^b}^{C} D_{t_{j(i)}^e}^{\alpha_{j(i)}}\left( \left\Vert
  \mbox{\boldmath{$\zeta$}}_{j(i)}(t)\right\Vert\right) = \left\{
  \begin{array}{lcl}
   \frac{1}{\Gamma\left(n_{j(i)}-\alpha_{j(i)}\right)} 
   \int\limits_{t_{j(i)}^b}^{t}\frac{\frac{d^{n_{j(i)}} }{d\tau^{n_{j(i)}}}\left
   \Vert \mbox{\boldmath{$\zeta$}}_{j(i)}(\tau)\right\Vert}
   {(t-\tau)^{\alpha_{j(i)}-n_{j(i)}+1}}d\tau & \mbox{for} & n_{j(i)}-1 <
   \alpha_{j(i)} < n_{j(i)} \\ \frac{d^{n_{j(i)}}}
   {d\left(t-t_{j(i)}^e\right)^{n_{j(i)}}}\left\Vert
   \mbox{\boldmath{$\zeta$}}_{j(i)}(t)\right\Vert & \mbox{for} &
   \alpha_{j(i)}=n_{j(i)}
  \end{array}
  \right.,
 \end{equation}
\end{widetext}
where $n_{j(i)}=\left[\alpha_{j(i)}\right]+1$ and $[\cdot]$ denotes an integer
part of a~real number, and
$
 _{t_{j(i)}^b}\!\!I_{t_{j(i)}^e}^{\beta_{j(i)}}\left( \left\Vert
  \mbox{\boldmath{$\zeta$}}_{j(i)}(t)\right\Vert\right)
$
is the Riemann-Liouville fractional integral
\begin{widetext}
 \begin{equation} \label{eq34}
 _{t_{j(i)}^b} I_{t_{j(i)}^e}^{\beta_{j(i)}}\left( \left\Vert
 \mbox{\boldmath{$\zeta$}}_{j(i)}(t)\right\Vert\right) =\left\{
 \begin{array}{lll}
  \frac{1}{\Gamma\left(\beta_{j(i)}\right)} \int\limits_{t_{j(i)}^b}^{t}
  \left\Vert \mbox{\boldmath{$\zeta$}}_{j(i)}(\tau)\right\Vert
  \left(t-\tau\right)^{\beta_{j(i)}-1} d\tau & \mbox{for} & \beta_{j(i)} \in
  {\cal R}^{+} \\
  \frac{1}{\left(\beta_{j(i)}-1\right)!} \int\limits_{t_{j(i)}^b}^{t}\left\Vert
  \mbox{\boldmath{$\zeta$}}_{j(i)}(\tau)\right\Vert
  \left(t-\tau\right)^{\beta_{j(i)}-1} d\tau & \mbox{for} & \beta_{j(i)}\in
  {\cal N} \end{array}\right.
 \end{equation}
\end{widetext}
and $\beta_{j(i)}=-\alpha_{j(i)}$. Eqn.~(\ref{eq31}) represents the form of the
repulsive force acting in the normal direction to the contacting surfaces.

Now we introduce additional definitions of forces operating in the tangent
plane. Here we define the normal force as
$\mathbf{N^{'}}_{j(i)}=\left[0,0,R_{\zeta\, j(i)}\right]$. According
to~\cite{Farkas} we define the friction force which is coupled between
torsion-sliding friction as
\begin{eqnarray} \label{eq36} 
 \mathbf{T}_{j(i)}^{'\, sli}&=&-\mu\left(\left\Vert
 \widetilde{\mathbf{u}}_{t\, j(i)}^{'\, lin}\right\Vert \right){\cal F}
 (\lambda_{j(i)})\nonumber \\ && N_{\zeta\, j(i)}\left[
 \begin{array}{c}
  \mbox{sign}\left(\widetilde{u}_{\xi\, j(i)}^{lin}-
  \widetilde{u}_{\xi\, j(i)}^{rot}\right)\\ \mbox{sign}\left(
  \widetilde{u}_{\eta\, j(i)}^{lin}-\widetilde{u}_{\eta\, j(i)}^{rot}\right)\\
  0
 \end{array}
 \right],
\end{eqnarray}
where the friction coefficient is
\begin{equation} \label{eq361}
 \mu\left(\left\Vert \widetilde{\mathbf{u}}_{t\, j(i)}^{'\, lin}\right\Vert
 \right)=\mu_{d}+\left(\mu_{s}-\mu_{d}\right)e^{-a\left\Vert
 \widetilde{\mathbf{u}}_{t\, j(i)}^{'\, lin}\right\Vert },
\end{equation}
where $a$ is a~numerical constant, $\mu_{s}$ and $\mu_{d}$ are static and
dynamic coefficients of friction. Moreover in formula~(\ref{eq36}) the function
${\cal F}\left(\lambda_{j(i)}\right)$ is defined according to~\cite{Farkas} as
\begin{widetext}
 \begin{equation} \label{eq362}
  {\cal F}(\lambda_{j(i)})=\left\{
  \begin{array}{lll}
   \frac{4}{3}\frac{\left(\lambda_{j(i)}^{2}+1\right)E\left(\lambda_{j(i)}
   \right)+\left(\lambda_{j(i)}^{2}-1\right)K\left(\lambda_{j(i)}\right)}
   {\pi\lambda_{j(i)}} & \mbox{for} & \lambda\leq1\\
   \frac{4}{3}\frac{\left(\lambda_{j(i)}^{2}+1\right)E\left(\frac{1}
   {\lambda_{j(i)}}\right)-\left(\lambda_{j(i)}^{2}-1\right)K\left(
   \frac{1}{\lambda_{j(i)}}\right)}{\pi} & \mbox{for} & \lambda>1
  \end{array}
  \right.,
 \end{equation}
\end{widetext}
where $K(\lambda_{j(i)})$ and $E(\lambda_{j(i)})$ are the complete elliptic
integral functions of the first and the second kind, $\lambda_{j(i)}$ is the
dimensionless quantity defined as
\begin{equation} \label{eq363}
 \lambda_{j(i)} = \frac{\left\Vert \widetilde{\mathbf{u}}_{t\, j(i)}^{'\, lin}
 -\widetilde{\mathbf{u}}_{t\, j(i)}^{'\, rot}\right\Vert }{\frac{1}{2}\left\Vert
 \mbox{\boldmath{$\eta$}}_{j(i)}\right\Vert \left|\omega_{\zeta\, i}-
 \omega_{\zeta\, j(i)}\right|}.
\end{equation}
The limiting values of the function ${\cal F}\left(\lambda_{j(i)}\right)$ are
${\cal F}(0)=0$ for torsion without sliding and
$
 \lim\limits _{\lambda_{j(i)}\rightarrow\infty}{\cal F}
 \left(\lambda_{j(i)}\right)=1
$
for sliding without torsion.

According to~\cite{Zhang1} we define the rolling friction force as
\begin{widetext}
 \begin{eqnarray} \label{eq37} 
  \mathbf{T}_{j(i)}^{'\, rol} &=& \frac{1}{\frac{1}{m_i}+\frac{1}{m_{j(i)}}+
  \frac{\left(\left\Vert \mbox{\boldmath{$\widetilde{\zeta}$}}_{j(i)}
  \right\Vert\right)^2}{{\cal J}_i}+\frac{\left(\left\Vert 
  \mbox{\boldmath{$\widetilde{\widetilde{\zeta}}$}}_{j(i)}\right\Vert\right)^2}
  {{\cal J}_{j(i)}}} \nonumber \\
  && \left(\frac{1}{{\cal J}_i}\mathbf{ss}_{i,j(i)}^{'}\times \mathbf{N}_{j(i)}
  \times\mathbf{s}_{i,j(i)}^{'}- \frac{1}{{\cal J}_{j(i)}}
  \mathbf{ss}_{j(i),i}^{'}\times \mathbf{N}_{j(i)}\times\mathbf{s}_{j(i),i}^{'}
  + \mathbf{A}_{j(i)}\right),
 \end{eqnarray}
\end{widetext}
where
\begin{equation} \label{eq3711}
 \mathbf{ss}_{i,j(i)}^{'}= \left[ \left\Vert \mbox{\boldmath{$\eta$}}_{j(i)}
 \right\Vert \mbox{sign}\left(\widetilde{u}_{\eta\,j(i)}^{rot}\right),
 \left\Vert \mbox{\boldmath{$\eta$}}_{j(i)}\right\Vert \mbox{sign}
 \left(\widetilde{u}_{\xi\,j(i)}^{rot}\right), 0\right]
\end{equation}
and
\begin{widetext}
 \begin{eqnarray} \label{eq371}
  \mathbf{A}_{j(i)}&=&\frac{1}{m_i}\sum\limits_{l(i)}\mathbf{e}_{j(i)} \cdot
  \mathbf{F}_{l(i)} - \frac{1}{m_{j(i)}} \sum\limits_{l(j(i))}
  \mathbf{e}_{j(i)} \cdot \mathbf{F}_{l(j(i))} - \frac{1}{{\cal J}_i}
  \sum\limits_{l(i)}\left( \mathbf{e}_{j(i)} \cdot \mathbf{M}_{l(i)} \right)
  \times \mathbf{s}_{i,j(i)}^{'} \nonumber \\ &&- \frac{1}{{\cal J}_{j(i)}}
  \sum\limits_{l(j(i))}\left( \mathbf{e}_{j(i)} \cdot \mathbf{M}_{l(j(i))}
  \right) \times \mathbf{s}_{j(i),i}^{'} - \mbox{\boldmath{$\omega$}}_{i}
  \times \frac{d\,\mathbf{s}_{i,j(i)}^{'}}{d\,t} -
  \mbox{\boldmath{$\omega$}}_{j(i)} \times
  \frac{d\,\mathbf{s}_{j(i),i}^{'}}{d\,t}.
 \end{eqnarray}
\end{widetext} 

The above expressions are necessary for the definitions of some collisional torques.
Therefore we have the collisional torque operating from particle $i$ to particle
$j(i)$ as
\begin{equation} \label{eq38}
 \mathbf{M}_{i,j(i)}^{coll}=\left\{
 \begin{array}{lll}
  \mathbf{0} & \mbox{for} & \left\Vert 
  \widetilde{\mathbf{u}}_{t\, j(i)}^{'\, lin}\right\Vert =\left\Vert
  \widetilde{\mathbf{u}}_{t\, j(i)}^{'\, rot}\right\Vert =0\\
  \mathbf{M}_{i,j(i)}^{rol} & \mbox{for} & \left\Vert
  \widetilde{\mathbf{u}}_{t\, j(i)}^{'\, lin}-
  \widetilde{\mathbf{u}}_{t\, j(i)}^{'\, rot}\right\Vert =0\\
  \mathbf{M}_{i,j(i)}^{sli} & \mbox{for} & \left\Vert
  \widetilde{\mathbf{u}}_{t\, j(i)}^{'\, lin}-
  \widetilde{\mathbf{u}}_{t\, j(i)}^{'\, rot}\right\Vert >0
 \end{array}
 \right.,
\end{equation}
where $\mathbf{M}_{i,j(i)}^{sli}$ is the coupled \ torsion-sliding \ torque,
$\mathbf{M}_{i,j(i)}^{rol}$ represents the \ coupled \ torsion-rolling torque. Note
that transition from the local system of coordinates to the global ones reads
\begin{eqnarray} \label{eq40}
 && \mathbf{M}_{i,j(i)}^{sli}=\mathbf{e}_{j(i)}^{T}\cdot\left(
 \mathbf{M}_{i,j(i)}^{'\, sli}+\mathbf{M}_{i,j(i)}^{'\, tor}\right),\nonumber \\
 && \mathbf{M}_{i,j(i)}^{rol}= \mathbf{e}_{j(i)}^{T}\cdot \left(
 \mathbf{M}_{i,j(i)}^{'\, rol} + \mathbf{M}_{i,j(i)}^{'\, tor} \right).
\end{eqnarray}
We define the torsion torque as
$\mathbf{M}_{i,j(i)}^{'\, tor}=\left[0,0,M_{\zeta\, i,j(i)}^{tor}\right]^{T}$
and according to~\cite{Farkas} we obtain
\begin{eqnarray} \label{eq42}
 M_{\zeta \, i,j(i)}^{tor} &=& - \frac{1}{2} {\cal T}\left(\lambda_{j(i)}\right)
 \left\Vert \mbox{\boldmath{$\eta$}}_{j(i)}\right\Vert \mu\left(\left\Vert
 \widetilde{\mathbf{u}}_{t\,j(i)}^{'\, lin} \right\Vert\right) \nonumber \\ &&
 N_{\zeta \,j(i)} \mbox{sign}\left(\omega_{\zeta\,i}-
 \omega_{\zeta\,j(i)}\right),
\end{eqnarray}
where the function ${\cal T}\left(\lambda_{j(i)}\right)$ reads
\begin{widetext}
 \begin{equation} \label{eq43}
  {\cal T}\left(\lambda_{j(i)}\right)=\left\{
  \begin{array}{lll}
   \frac{4}{9}\frac{\left(4-2\lambda_{j(i)}^{2}\right)E\left(\lambda_{j(i)}
   \right)+\left(\lambda_{j(i)}^{2}-1\right)K\left(\lambda_{j(i)}\right)}{\pi} &
   \mbox{for} & \lambda_{j(i)}\leq1\\
   \frac{4}{9}\frac{\left(4-2\lambda_{j(i)}^{2}\right)E\left(\frac{1}
   {\lambda_{j(i)}}\right)+\left(2\lambda_{j(i)}^{2}-5+\frac{3}
   {\lambda_{j(i)}^{2}}\right)K\left(\frac{1}{\lambda_{j(i)}}\right)}
   {\pi\lambda_{j(i)}} & \mbox{for} & \lambda_{j(i)}>1
  \end{array}
  \right..
 \end{equation}
\end{widetext}
The limiting values of the function ${\cal T}\left(\lambda_{j(i)}\right)$ are
${\cal T}(0)=\frac{2}{3}$ for torsion without sliding and
$
 \lim\limits _{\lambda_{j(i)}\rightarrow\infty}{\cal T}\left(\lambda_{j(i)}
 \right)=0
$
for sliding without torsion. Moreover, we introduce the sliding torque as
\begin{equation} \label{eq41}
 \mathbf{M}_{i,j(i)}^{'\, sli}=-\mathbf{s}_{i,j(i)}^{'}\times
 \mathbf{T}_{j(i)}^{'\, sli}.
\end{equation}

Using an idea included in~\cite{Zhang1} we determine the rolling torque as
\begin{equation} \label{eq44}
 \mathbf{M}_{i,j(i)}^{'\, rol}=-\mathbf{s}_{i,j(i)}^{'}\times
 \mathbf{T}_{j(i)}^{'\, rol}+\mathbf{ss}_{i,j(i)}^{'}\times
 \mathbf{N}_{j(i)}^{'},
\end{equation}
where $\mathbf{ss}_{i,j(i)}^{'}\times\mathbf{N}_{j(i)}^{'}$ is the torque
created on the penetration width~(\ref{eq05}). As noted in~\cite{Zhang1},
the torque $\mathbf{ss}_{i,j(i)}^{'}\times\mathbf{N}_{j(i)}^{'}$ exists
because the contact between two particles is not a~single point but, due to
deformation of both bodies, is a~finite area.

Summarising this subsection we determined a~full description of the forces and
torques occurring in a~collision. We neglect here a~mathematical description
of the collisional force $\mathbf{P}_{j(i)}^{b\, coll}$ and torque
$\mathbf{M}_{i,j(i)}^{b\, coll}$ acting between the particle-wall because one
can easily produce these formulae taking into account
$\mathbf{\dot{x}}_{j(i)}=\mathbf{0}$,
$\mbox{\boldmath{$\omega$}}_{j(i)}=\mathbf{0}$, etc. in the above expressions.
More details concerning particle-wall interaction can be found in~\cite{Kondic}.

\section{Solution procedure}
Throughout this section we will show how to handle the system of ordinary
differential equations~(\ref{eq21}) and (\ref{eq22}) in order to simulate the
dynamics of multiparticle collisions. The above system is mathematically
complex, and therefore requires a~numerical approach. An accurate solution to
this problem
was obtained by integrating the system of ordinary differential
equations~(\ref{eq21}) for particles moving individually by using Numerical
Recipe routines~\cite{Press}. Tracing the motion of individual particles over
time
we need to detect particle collisions in order to take into account collisional
forces and torques in the system of differential equations. Using results
presented in~\cite{Allen,Iwai,Schinner} we have chosen the linked cell method
to detect a~collision. 

Note that during particle collisions we need to solve the system~(\ref{eq22})
where the fractional interaction law~(\ref{eq31}) occurs. In this case we have
a~system of ordinary differential equations with a~mixture of operators: the
integer derivative of maximal order equals two, the fractional integral of
order $-\alpha_{j(i)}$ and the fractional derivative of order $\alpha_{j(i)}$.
Using fractional calculus~\cite{Oldham,Samko} we will present discrete
forms of the fractional operators which are suitable in our algorithm. Let us
consider the duration of a~collision over time
$t\in\left\langle t_{0},t_{nt}\right\rangle $ where $t_{0}$ represents the time
when the collision starts and $t_{nt}$ is the time when the collision ends, $nt$
denotes the division of the collisional time $t$ into several time steps.
Thus we obtain: $h=\frac{t_{nt}-t_{0}}{nt},$ $t_{l}=t_{0}+lh,$ for
$l=0,\ldots,nt$. If a~function $f(t)$ is constant within the step $h$ then the
discrete form of the Caputo fractional derivative~(\ref{eq33}) becomes
\begin{widetext}
 \begin{equation} \label{eq45}
  _{t_{0}}^{C}D_{t_{nt}}^{\alpha}f\left(t\right)=\frac{1}{\Gamma(n-\alpha+1)}
  \left[A_{1}\left(t_{nt}-t_{0}\right)^{n-\alpha}+\sum\limits _{l=2}^{nt}\left(
  A_{l}-A_{l-1}\right)\left(t_{nt}-t_{l-1}\right)^{n-\alpha}\right],
 \end{equation}
\end{widetext}
where $\alpha\in{\cal R}^{+}$, $n=[\alpha]+1$ and $[\cdot]$ denotes an integer
part of a~real number, $A_{l}=f^{(n)}\left(t_{l}\right)$ where $f^{(n)}$ is the
derivative of integer order $n$. Note that in formula~(\ref{eq45}) $f(t)$
denotes the overlap~(\ref{eq04}). Taking the above assumptions into account we
obtain the discrete form of the Riemann-Liouville fractional
integral~(\ref{eq34}) as
\begin{widetext}
 \begin{equation} \label{eq46}
  _{t_{0}}I_{t_{nt}}^{\beta}f\left(t\right)=\frac{1}{\Gamma(\beta+1)}\left[
  B_{1}\left(t_{nt}-t_{0}\right)^{\beta}+\sum\limits _{l=2}^{nt}\left(
  B_{l}-B_{l-1}\right)\left(t_{nt}-t_{l-1}\right)^{\beta}\right],
 \end{equation}
\end{widetext}
where $\beta\in{\cal R}^{+}$and $B_{l}=f\left(t_{l}\right)$. The Discrete forms
of the fractional operators makes it possibile to integrate the
system~(\ref{eq22}) by using any predictor-corrector procedure~\cite{Press}
with correction of the time step $h$. The correction of the time step provides
measures that allow us to determine the begin time when particles enter into
a~collision and the end time of particle collisions. It should be noted that the
begin and end times are determined by several definitions presented in the
previous section.

Taking formulae~(\ref{eq28}), (\ref{eq38}) into account in calculations of
particle contacts we need to find an accurate time needed to detect the
switching between torsion-sliding, sliding and rolling processes. As described
in the paper~\cite{Zhang1}, a~simple way to calculate the switching time is to
use a~linear approximation method.

Next we consider a~problem occurring in the calculations of friction
forces~(\ref{eq36}), (\ref{eq37}) and the torsional torque~(\ref{eq42}). When
the relative velocity at the contact point changes from negative to positive or
from positive to negative, it indicates that the signum function
$\mbox{sign}(x)$ changes sign very fast in above expressions. This is not
desirable as it influences the stability and convergence of the numerical
calculations in a~significant way. Therefore we modified the signum function
introducing
\begin{equation} \label{eq47}
 \widetilde{\mbox{sign}}(x)=\left\{
 \begin{array}{lll}
  -1 & \mbox{for} & x\leq-\epsilon_{2}\\
  \frac{1}{\epsilon_{2}-\epsilon_{1}}x+\frac{\epsilon_{1}}
  {\epsilon_{2}-\epsilon_{1}} & \mbox{for} & -\epsilon_{2}\leq x\leq-
  \epsilon_{1}\\
  0 & \mbox{for} & -\epsilon_{1}\leq x\leq\epsilon_{1}\\
  \frac{1}{\epsilon_{2}-\epsilon_{1}}x-\frac{\epsilon_{1}}
  {\epsilon_{2}-\epsilon_{1}} & \mbox{for} & \epsilon_{1}\leq x\leq\epsilon_{2}
  \\ 1 & \mbox{for} & x\geq\epsilon_{2}
 \end{array}
 \right.,
\end{equation}
where $x$ is the actual value registered during a~contact (the relative
velocity), $\epsilon_{1}$, $\epsilon_{2}$ are numerical coefficients. This
function is robust for $x\rightarrow0$ and gives a~satisfactory result.

\section{Results and their analysis}
\begin{figure*}[ht]
 \begin{center}
  \includegraphics[width=1.0\textwidth,keepaspectratio]{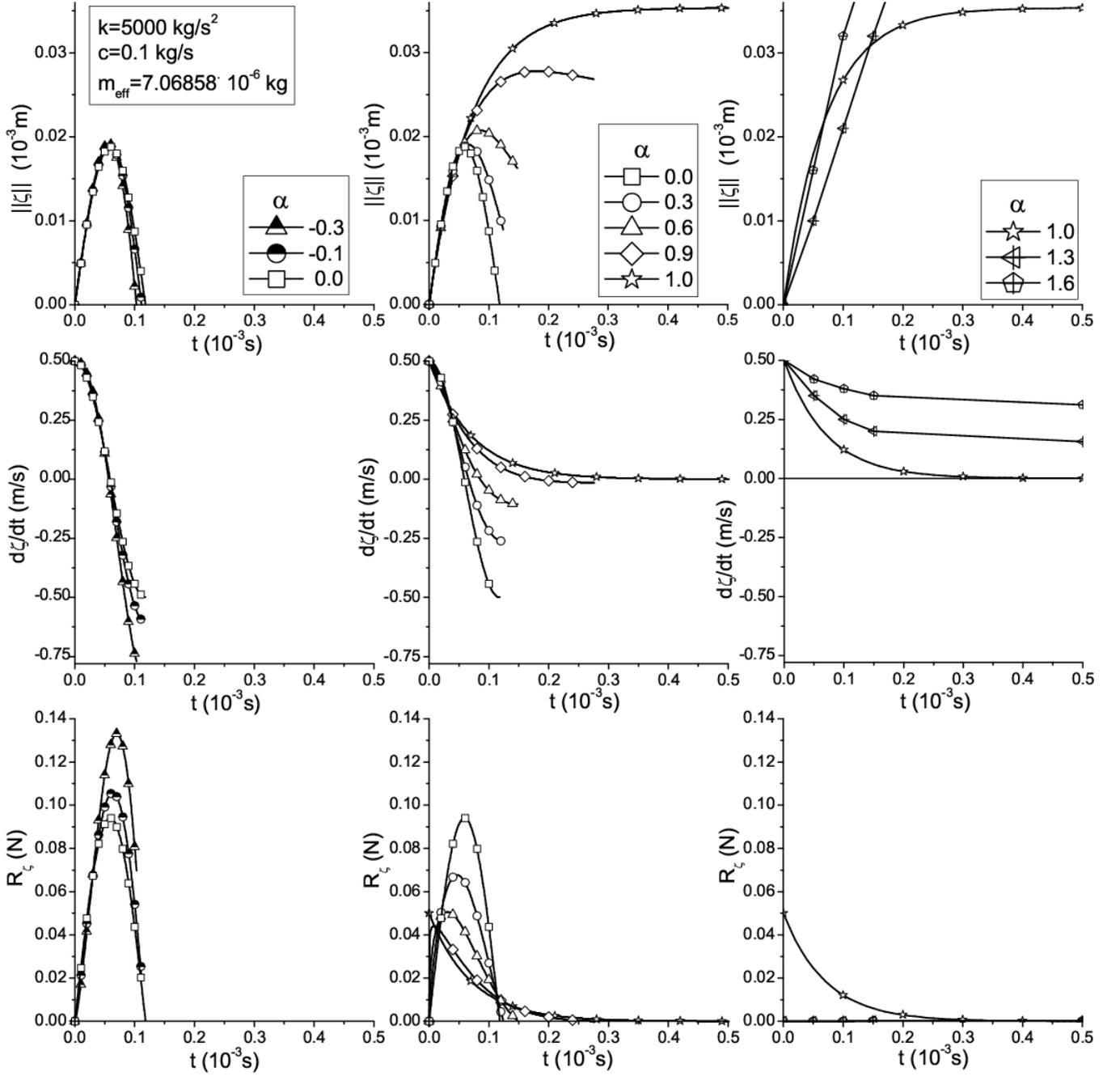}
 \end{center}
 \caption{Behaviour of the overlap (top), relative velocity (middle) and force
          (bottom) over time for the fractional interaction law.\label{fig02}}
\end{figure*}

To illustrate the benefits of the fractional interaction law~(\ref{eq31})
in the dynamics of arbitrary multiparticle collisions, first we will demonstrate
how this law operates in simple cases connected with a~one dimensional
problem. First we will simulate a~central collision between two particles.
Fig.~\ref{fig02} shows the dynamics of a~two-particle collision, which is
represented by some variations in the overlap $\left\Vert \zeta\right\Vert $
(\ref{eq04}), the linear relative velocity $\dot{\zeta}=u_{\zeta}^{'\, lin}$
(\ref{eq0013}) and the repulsive force $R_{\zeta}$ (\ref{eq31}) over time for
different levels of the conversion degree $\alpha.$ Here we neglect the index
$j(i)$ because only two particles collide. Moreover, all vectors are converted
to scalar values when a~one dimensional problem is considered. For the figure we
have $m_{eff}=\frac{m_{1}m_{2}}{m_{1}+m_{2}}=7.06858\cdot10^{-6}\, kg$,
$r_{1}=r_{2}=3\cdot10^{-3}\, m$, $k=5000\,\frac{kg}{s^{2}}$, $c=0.1\, kg/s$. The
initial relative velocity is set at $\dot{\zeta}=0.5\,\frac{m}{s}$ and three
groups of variations in the conversion degree are taken into account. The first
group is for $\alpha<0$ (left column), the second is for $0\leq\alpha\leq1$
(middle column) and the third represents $\alpha>1$ (right column).
Within the range $0\leq\alpha\leq1$ we observe that collisional time $t_{c}$
increases when $\alpha$ is increased. It should be noted that the collisional
time is registered when the repulsive force $R_{\zeta}$ reaches zero, as
presented in several definitions in the previous sections. Therefore, the
overlap
$||\zeta||$ has some values at the time when a~collision ends and deformations
of the particle surfaces are noted. Analysing the behaviour of the relative
velocity $\dot{\zeta}$ over time we notice that this velocity changes direction
for small values of $\alpha$, which means thatparticle rebounds dominate. When
$\alpha$ increases we can see that the relative velocity tends to zero, which
means that particles stick together. In other words, if $\alpha=0$, no viscous
term in
Eqn.~(\ref{eq31}) may occur and all the impact energy must be due to elasticity.
In this case the overlap reaches zero at the same time as the repulsive force
reaches zero. If $\alpha=1$, on the other hand, the impact energy is
transfered through the viscous term. \\
Extending our considerations for $\alpha>1$ we observe (right column on
Fig.~\ref{fig02}) that the repulsive force is not generated and tends to zero
for $t_{c}\rightarrow\infty$, and therefore the overlap increases to high and
unrealistic values. Moreover, the relative velocity does not change direction
and particles undergo the next time steps of the calculations. According to
definition~6, presented in the previous section, the fragmentation of particles
or permanent cohesion of particles is a~direct result of the plastic flow of
their contacting surfaces. The contacting surfaces are destroyed because
deformations of contacting particles become
sufficiently large so as to exceed the elastic limit of the materials, and we
noticed particle clusterisations. This process is observed experimentally
in~\cite{Krupp,Rimai} and may be modelled by the fractional interaction
law~(\ref{eq31}).\\
Next we considered the behaviour of the overlap, relative velocity and repulsive
force for $\alpha<0$ (left column on Fig.~\ref{fig02})). Larger negative values
of the conversion degree $\alpha$ decreases the collisional time. The relative
velocity changes direction but at the end time reaches larger absolute values
in comparison to the initial relative velocity. As this is unrealistic all the
solutions for $\alpha<0$ are not taken into account. The aim of this example is
to show the power of fractional calculus where more solutions are obtained
in comparison to classical differential and integral operators having integer
order. However, we need to choose which solutions obtained by fractional
calculus are suitable physically.

In Fig.~\ref{eq03} we constructed several mappings for the relative
velocity-overlap (left), force-overlap (middle) and force-relative velocity
(right) where $\alpha$ changes from negative to positive values.
\begin{figure*}[ht]
 \begin{center}
  \includegraphics[width=1.0\textwidth,keepaspectratio]{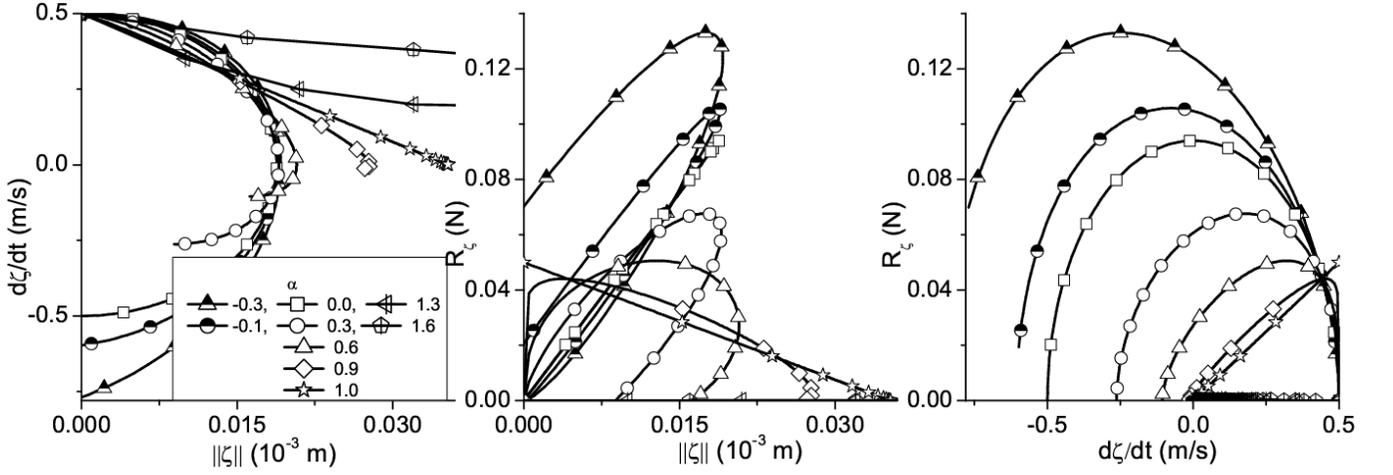}
 \end{center}
 \caption{Mapping the relative velocity-overlap (left), force-overlap (middle)
          and force-relative velocity (right) for the fractional interaction
	  law. \label{fig03}}
\end{figure*}
Analysing these mappings we found a~set of criteria necessary to predict
different states of particle collisions included in definitions in the previous
section. It should be noted that small positive values of $\alpha$ predict
particle rebounds when particle deformations are practically neglected. When
$\alpha$ tends to unity we also observe particle rebounds but particle
deformations are visible and more energy is dissipated. As indicated in the left
chart in Fig.~\ref{eq03}, when $\alpha$ is above unity the repulsive force is not
generated and this indicates instability in particle collisions. This
instability takes the form of particle fragmentation or permanent clusterisation of
particles after the collision. Therefore the conversion degree $\alpha$ is
a~ratio of the impact energy over the specific energy needed for the destruction of
particle surfaces. This assumption should be validated experimentally, and
this is the aim of our future investigations. Note that when the physical
properties of colliding granular materials and the impact energy are fixed we
still observe different values of energy dissipation after the collision. This
can be easily seen when we compare the particle collisions for smooth surfaces of
particles and for rough ones. The fractional interaction law can simulate this
because the conversion degree $\alpha$ can change. 

\begin{figure}[ht]
 \begin{center}
  \includegraphics[width=0.90\columnwidth,keepaspectratio]{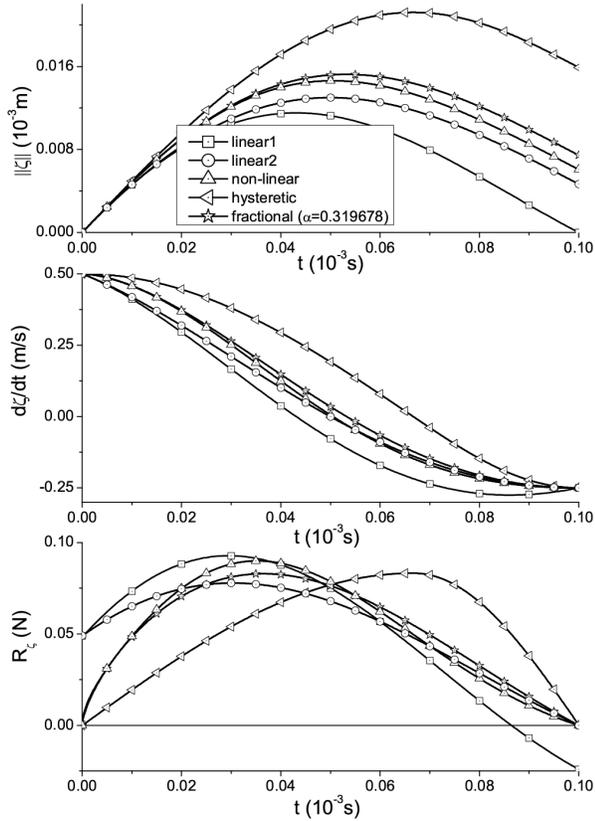}
 \end{center}
 \caption{Comparison of the overlap (top), relative velocity (middle) and force
          (bottom) over time for different interaction laws.\label{fig04}}
\end{figure}

In order to compare the fractional interaction law with other interaction laws,
changes over time of the overlap, the relative velocity and the repulsive force
for two-particle collision were presented. We assumed parameters of colliding
particles to be $r_{1}=r_{2}=3\cdot10^{-3}\, m$, $m_{eff}=7.06858\cdot10^{-6}\, kg$,
$\dot{\zeta}=0.5\,\frac{m}{s}$. Moreover, we assumed the collision time between
two colliding bodies the $t_{c}=10^{-4}\, s$ and the restitution coefficient
$e_{r}=0.5$. These assumptions are necessary to calculate the set of coefficients
required by different interaction laws, depending on the type of interaction law
chosen. In Table~\ref{tab01} we list all the coefficients.
\begin{table}
 \caption{Coefficients for colliding particle surfaces being dependent on the
          interaction law used. \label{tab01}}
 \begin{ruledtabular}
  \begin{tabular}{ll}
   law & coefficients \\ \hline \\
   linear1 & \( k_n=7316\, \frac{kg}{s^2}\), \( c_n=0.0979\, \frac{kg}{s}\) \\
   linear2 & \( k_n=5225\, \frac{kg}{s^2}\), \( c_n=0.0981\, \frac{kg}{s}\) \\
   non-linear & \( \widetilde{k}=1392000\, \frac{kg}{s^2\sqrt{m}}\), 
    \( \widetilde{c}=33.885\, \frac{kg}{s\sqrt{m}}\) \\
   hysteretic & \( k_1=3924\, \frac{kg}{s^2}\), \(k_2=15697\, \frac{kg}{s^2}\)\\
   fractional & \( k=5225\, \frac{kg}{s^2}\), \( c=0.297\, \frac{kg}{s}\),
    \(\alpha=0.3197\)
  \end{tabular}
 \end{ruledtabular}
\end{table}
Some of the expressions applied to calculate the coefficients for
linear~\cite{Cundall} and hysteretic~\cite{Walton} laws can be found
in~\cite{Pourin}. The formulae of the coefficients used in the linear
interaction law assumed that at the end time of a~collision the overlap is zero.
In Table~\ref{tab01} the ``linear1'' represents the above case. We assumed
that the repulsive force reaches zero at the end time of a~collision. Thus we have
a~set of coefficients called ``linear2'' also used for the linear interaction
law. For the non-linear~\cite{Kuwabara} and fractional laws we performed
a~numerical test to find the values of coefficients which allow us to keep the
assumed collision time and the restitution coefficient in a~two-particle
contact. It should be noted that we obtained many sets of coefficients for the
fractional interaction law. Therefore for this law we establish the spring
coefficient which has the same value as for the linear interaction law.
Fig.~\ref{fig04} shows the behaviour of the overlap (top chart), the relative
velocity (middle chart) and the repulsive force (bottom chart) over time where
different interaction laws are taken into account.
Analysing this figure we can confirmed that the interaction law fulfilled our
assumptions concerning the collisional time and the restitution coefficient. It
should be noted that the repulsive force changes direction in the linear
interaction law (linear1) for the set of coefficients calculated under the formulae
found in~\cite{Pourin}. This shows a~deficiency in numerical calculations and
should be rejected. Some changes in the values of the above coefficients give
satisfactory results in the linear interaction law (linear2). However, the
repulsive force in the linear interaction law has a~value at the beginning time
which is independent on the set of coefficients used. This is also unrealistic
behaviour in the linear law. \\
Using different interaction laws we observed different overlaps at the end time
of a~collision. The highest value of overlap is for the hysteretic law and
decreases for the fractional through the non-linear to the linear one. Note
that for the fractional law we can find another set of coefficients in order to
fulfil our assumptions and to obtain another value of the overlap at the end
time of collision.\\
When we determine all parameters necessary to describe the dynamics of particle
impacts then we obtain some values of the collisional time and the restitution
coefficient for this case. However when we still keep the above parameters and
increase or decrease the surface roughness of colliding particles then we obtain
values of the collisional time and the restitution coefficient differing in
comparison to the previous values. As we did not change physical properties of
this granular material therefore we have to save the steady value of the spring
coefficient in all interaction laws. Changing only the damping coefficient in
the linear and non-linear laws and the unloading slope $k_{2}$ in the hysteretic
law we do not have any guarantee that we will obtain accurate values of the
collisional time and the restitution coefficient reflecting the above cases. This is
a~disadvantage of the well-known interaction laws. In the fractional interaction law
we have an additional parameter called the conversion degree $\alpha$ which causes
some changes in the collisional time and the restitution coefficient.
However, this requires some experimental data involving the impact
dynamics of smooth and rough particles. These data will provide measures that
allow some links to be made between the experiment and the coefficients of the
fractional law.

\begin{figure}[ht]
 \begin{center}
  \includegraphics[width=1.0\columnwidth,keepaspectratio]{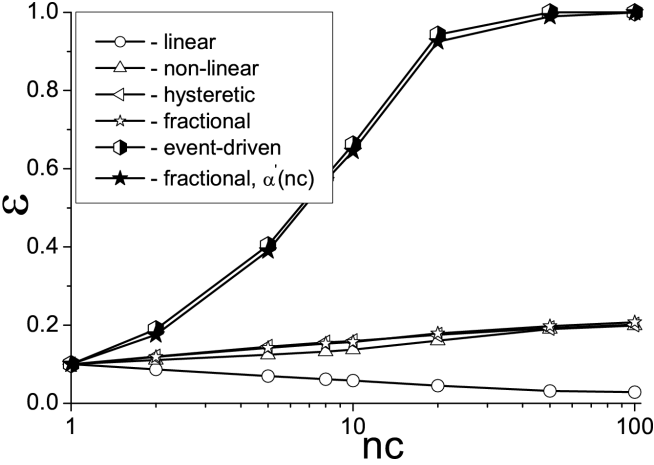}
 \end{center}
 \caption{Energy dissipation during multiparticle collisions for different
          interaction laws.\label{fig05}}
\end{figure}

In order to verify the validity of the interaction laws for multiparticle
collisions, the energies dissipated at each contact were compared. Here we
introduce a~measure of energy dissipation during multiparticle collisions which
is the ratio of the kinetic energy evaluated in time over the initial kinetic
energy. We define the total ratio of energy lost through multiparticle collisions as
\begin{equation} \label{eq48}
 \varepsilon=1-\frac{\sum\limits _{i=1}^{nc}m_{i}\dot{x}_{i}^{2}}
 {\sum\limits _{i=1}^{nc}m_{i}^{0}\left(\dot{x}_{i}^{0}\right)^{2}},
\end{equation}
where the superscript $0$ refers to the initial kinetic energy examined at time
$t=0\, s$ and $nc$ is the total number of colliding particles. \\
We used a~set of particles $np$ vertically stacked over a~bottom plate as shown
in~\cite{Luding,Pourin}. We assumed the following conditions $r_{i}=0.0015\, m$,
$m_{i}=1.41\cdot10^{-5}\, kg$, $\dot{x}_{i}=-0.5\,\frac{m}{s}$, for
$i=1,\ldots,np$. Gravity is set at zero. Taking into account the results presented
by~\cite{Luding} we calculated the energy dissipation as a~function of the
number of considered particles $np$, which becomes the number of colliding
particles $nc$ when at the begin time of the collision the distances between spheres
equal zero $l_{j}^{0}=0\, m$, for $j=1,\ldots,nc$. Note that $j=1$ represents
a~collision between the first particle and the bottom plate and $j=nc$ is
a~collision between the topmost particles. We also assume the collisional time
between two colliding bodies $t_{c}=10^{-4}\, s$ and the restitution coefficient
$e_{r}=0.945$. These assumptions are necessary to calculate some coefficients
depending on the type of interaction law chosen. The coefficients represent
a~collision between two particles or between a~particle and the bottom plate,
where the plate mass is infinite. \\
Fig.~\ref{fig05} shows the energy dissipation as being dependent on the number of
collisions $nc$ for different interaction laws used in the molecular dynamics
method and also in the event driven method~\cite{Allen,McNamara}.
For linear, non-linear and hysteretic laws we noted the same dependencies as
in~\cite{Luding,Pourin}. This means that the ``detachment'' effect occurs.
First, we considered the fractional interaction law for a~steady value of the
conversion degree $\alpha_{j(i)}=0.0258$, for all binary collisions. In this
case we obtained similar results for the hysteretic and fractional interaction
laws. Thus the ``detachment'' effect also occurs in the fractional interaction
law for the steady value of $\alpha_{j(i)}$. As written in~\cite{Luding} the
kinetic energy obtained from the event driven technique is dissipated totally
for $nc\cdot(1-e_{r})$ large. It should be noted that the basic interaction laws
are valid for a~two-particle collisions which are completely independent of other
collisions. However, in multiparticle collisions we need to include mutual
dependencies between several binary collisions. Taking this fact into account, we
can obtain satisfactory results when the conversion degree
$\alpha_{j(i)}$ changes in relation to the number of colliding particles. This was
explained more precisely in~\cite{Leszczynski}. Therefore we propose
$\alpha^{'}(nc)\sim1+\exp(-nc)$ in order to keep a~qualitative agreement with
the event driven method. It should be noted that we cannot estimate correctly
$\alpha^{'}(nc)$ by direct comparison with the event driven technique. We
require experimental data involving multiparticle collisions. This data will
provide measures that allow some links to be made between several coefficients
in the fractional interaction law and the experiment.

The last example simulates the dynamics of five particles in three dimensional
space for two values of the parameter $ \alpha$. The first value $ \alpha=0.01$
indicates the strong repulsive state, i.e. particles rebound almost without
dissipation of their energy. The second one for $ \alpha=0.97 $ represents the
weak repulsive state where most of the impact energy is converted into
material viscoelastcity. In the real behaviour of granular materials we can
easily observe such states, when we consider the collisions for contacting
particles with smooth surfaces and for rough ones. For this simulation we assumed
the following conditions $ r_1=0.02 \; m$, $r_2=0.01 \; m$, $r_3=0.007 \; m$,
$r_4=0.005 \; m$, $r_5=0.009 \; m$,
$\varrho_1=\varrho_4=2000 \; \frac{kg}{m^3}$,
$\varrho_2=\varrho_3=\varrho_5=1000 \; \frac{kg}{m^3}$,
$\mathbf{x}_1=\left[0.0,0.1,0.23\right]\; m$,
$\mathbf{x}_2=\left[0.001,0.125,0.205\right]\; m$,
$\mathbf{x}_3=\left[-0.002,0.090,0.198\right]\; m$,
$\mathbf{x}_4=\left[-0.004,0.120,0.186\right]\; m$,
$\mathbf{x}_5=\left[-0.001,0.1,0.18\right]\; m$. Moreover, we consider
a~situation where a~particle with an initial linear velocity
$\mathbf{u}_1=\left[0,0,-5\right]\; \frac{m}{s}$ collides at different moments
in time with particles which initially do not move
($\mathbf{u}_j=\left[0,0,0\right]\; \frac{m}{s}$, for $j=2,\ldots,4$). Particles
do not rotate initially
($\mbox{\boldmath{$\omega$}}_i=\mathbf{0}\; \frac{1}{s}$),
gravity is set to zero and $k=1000 \; \frac{kg}{s^2}$, $c=1 \; \frac{kg}{s}$
for each pair of colliding particles. We also simplified values of the friction
coeffcients putting into Eq.~(\ref{eq361}) $a=0$ and $\mu_s=0.5$ for each
pair of  colliding particles.
\begin{figure}[ht]
 \begin{center}
  \includegraphics[width=0.9\columnwidth,keepaspectratio]{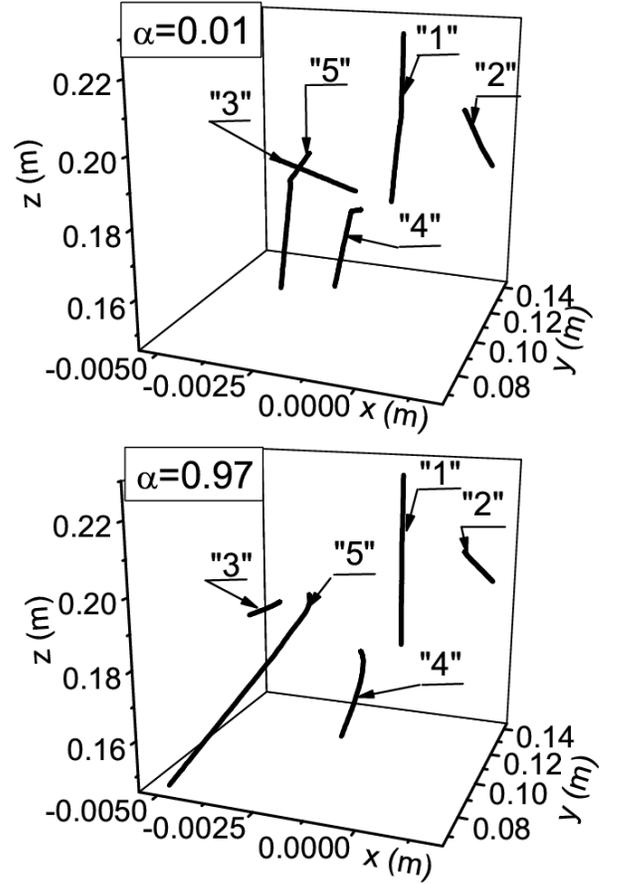}
 \end{center}
 \caption{Behaviour of particle trajectories depending on strong
          ($ \alpha=0.01 $) and weak ($ \alpha=0.97 $) repulsions.
	  \label{fig06}}
\end{figure}
Fig.~\ref{fig06} shows the trajectories of the mass centres of five particles in three
dimensional space for strong and weak repulsions as a~reaction to the impact
dynamics. The particles are numbered from "1" to "5".
\begin{figure*}[ht]
 \begin{center}
  \includegraphics[width=1.0\textwidth,keepaspectratio]{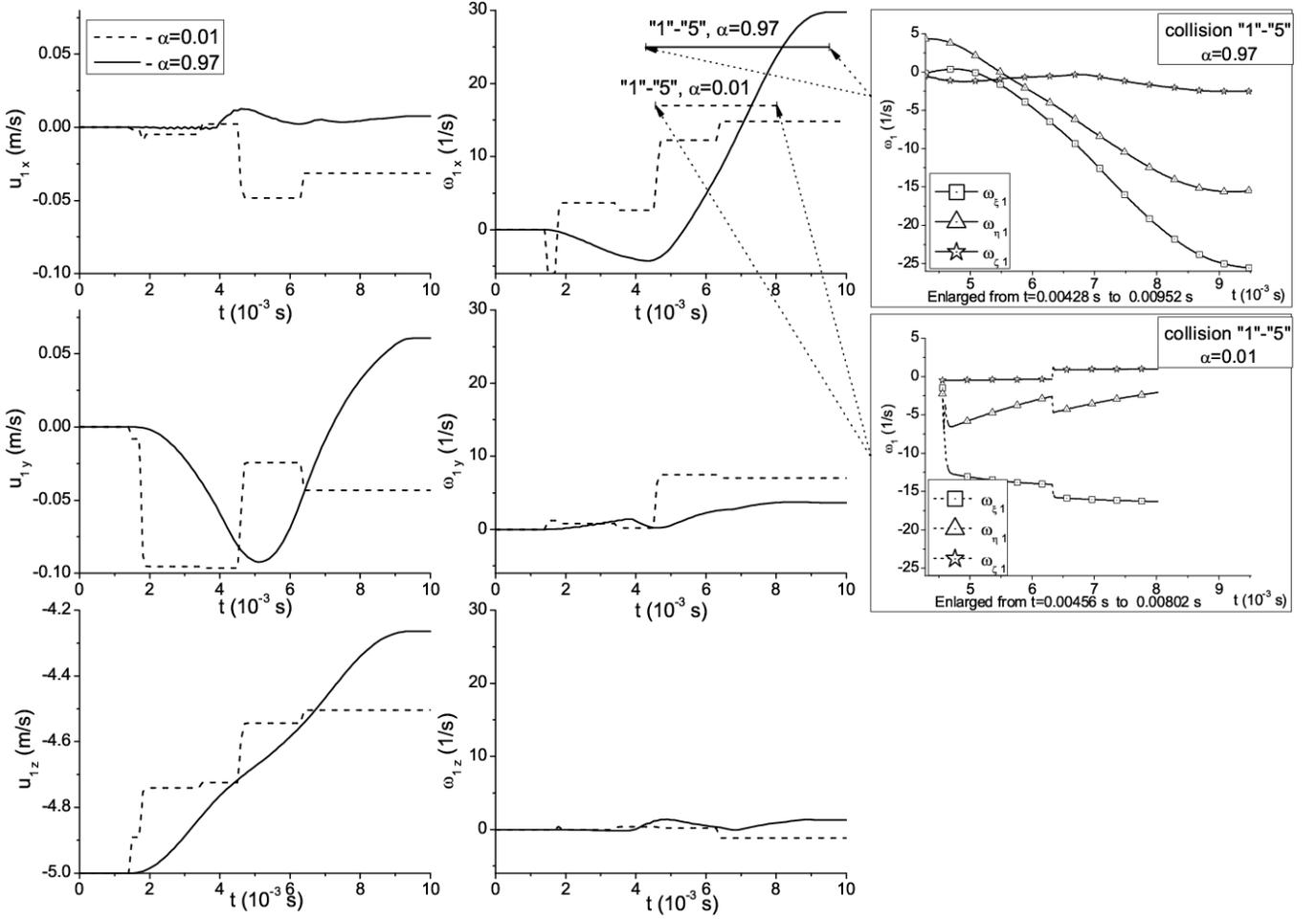}
 \end{center}
 \caption{Linear and angular velocities of particle "1" over time for
          strong ($ \alpha=0.01 $) and weak ($ \alpha=0.97 $) repulsions.
	  \label{fig07}}
\end{figure*}
This simulation does not reflect the real motion of particles because we
neglect external forces, i.e. the gravitational force. We can only show how the
fractional interaction law operates in the above conditions as being dependent
on the conversion degree $\alpha$. In the strong repulsive state ($\alpha=0.01$)
we observe linear particle trajectories. As $\alpha$ is increased and reaches
the weak repulsive state ($\alpha=0.97$) we noticed different particle
trajectories in comparison to the previous state. According to the results presented
in Fig.~\ref{fig02} we can say that duration over time of the repulsive force,
which is longer over time for higher values of $\alpha$, has a~significant
influence on the particle trajectories. \\
In order to explain more precisely what happens to particle trajectories in
strong and weak repulsive states, the velocities of one individual particle were
analysed. Fig.~\ref{fig07} shows in global coordinates $(x,y,z)$ the linear and
angular velocities of particle "1" over time. In this figure the dashed lines
represent particle velocities in the strong repulsive state, whereas continous
lines indicate the weak repulsive state. We can observe clear jumps in particle
velocities over time for the strong repulsive state. This is a~result of the
duration of a~collision determined by the collisonal time between a~pair of
contacting particles. In this state we can notice binary collisions because
several collisional times between the different pairs of contacting particles
have shortest values in comparison to their separation times, where particles
move individually. However, in the weak repulsive state we observe continous
changes in particle velocities without the distinction of any jumps. This means that
several collisional times between the pairs of contacting particles  overlap
each other. The binary collisions are not distinguished here. \\
Moreover, we analysed, in the local system of coordinates $(\xi,\eta,\zeta)$, the
angular velocities over time of particle "1", which collides with the particle
"5". In the strong repulsive state we observe smaller values of $\omega_{\xi}$
and $\omega_{\eta}$ (these velocities are angular velocities predicted in the
tangent plane as shown in Fig.~\ref{fig01}) in comparison to the weak repulsive
state. This means that torsion-sliding friction dominates in the strong repulsive
state, where binary collisions are noted. In the weak repulsive state we observe
that the angular velocities $\omega_{\xi}$ and $\omega_{\eta}$ have higher values
than in the strong repulsive state. Thus we expect the torsion-rolling state
between particles "1" and "5". However multiparticle collisions are noted in the
weak repulsive state.

\begin{figure}[ht]
 \begin{center}
  \includegraphics[width=1.0\columnwidth,keepaspectratio]{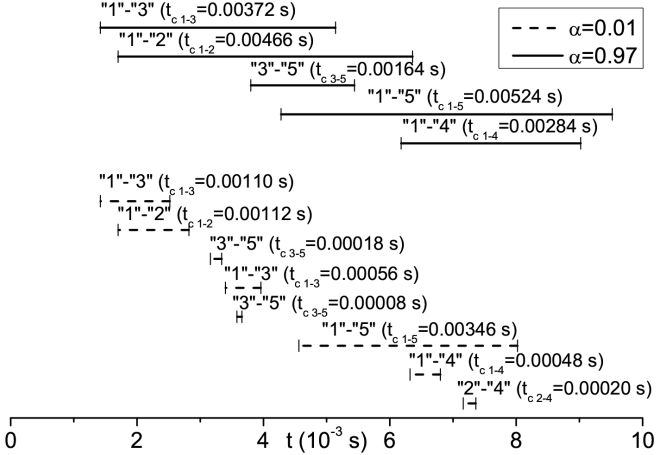}
 \end{center}
 \caption{Sequence of collisional times depending on strong ($ \alpha=0.01 $)
          and weak ($ \alpha=0.97 $) repulsions. \label{fig08}}
\end{figure}

In order to prove where binary or multiparticle collisions occur, some
distributions of collisional times over the duration time of calculations are
presented.
Fig.~\ref{fig08} presents the sequence of segments of collisional times over the
time of observation for strong and weak repulsive states.
Continous segments represent collisional times for weak repulsion whereas
strong repulsion is denoted by the dashed segments. Each segment representes one
binary collision between a~pair of contacting particles, i.e. "1"-"3" means the
collision between particle "1" and particle "3". Analysing this figure we
observe longer collisional times for the weak repulsive state in comparison to
the collisonal times for the strong repulsive state. Moreover the collisional
times ovelap in the weak repulsive state, therefore multiparticle collisions
occur. 

\section{Concluding remarks}
We used the molecular dynamics method to model the motion of individual
spherical particles in three-dimensional space. We introduced a~novel
mathematical description of this method which takes into account the division of the
collision process into an impact phase, contact phase and another phase formed
after the contact phase. We assumed that the impact phase and the phase formed
after the contact phase are infinitesimally short in time. We redefined the
collisional time so that it is predicted by the repulsive force-overlap path. On the
base of preliminary results~\cite{Leszczynski} we proposed an expression for the
repulsive force formulated under fractional calculus. The force can control the
energy dissipation and the collisional time for an individual particle colliding
with many other particles. In multiparticle collisions we included the
friction mechanism needed for the transition from coupled torsion-sliding
friction through rolling friction to static friction. Therefore our
model includes multiparticle collisions in arbitrary forms. Using the fractional
interaction law one can determine different states of particle repulsions, i.e.
strong and weak repulsive states. In the strong repulsive state binary collisions
dominate, and torsion-sliding friction is the main friction mechanism.
However, within the multiparticle collisions rolling friction is observed to be
much stronger. \\
The presented numerical results can be used to realistically model the impact
dynamics of an individual particle in a~group of colliding particles. In order
to tune the model coefficients we require experimental data involving
multiparticle collisions. This data provides measures that allow some links
to be made between several coefficients in the fractional interaction law and the
experiment.

\begin{acknowledgments}
 This work was supported by the \ State \ Committee for Scientific Research (KBN)
 under \ the \ grant \ 4~T10B~049~25.
\end{acknowledgments}


\begin{thebibliography}{10}
 \bibitem{Allen}        M.P. Allen and D.J. Tildesley, \emph{Computer
                        simulation of liquids} (Oxford University Press, Oxford,
		        1987).
 \bibitem{Balzer}       G. Balzer, Powder Technol. \textbf{113}, 299 (2000).
 \bibitem{Clement0}     E. Cl\'ement \emph{et al}., Int. J. Modern Phys. B
                        \textbf{7}, 1807 (1993).
 \bibitem{Cundall}      P.A. Cundall and O.D.L. Strack, Geotechnique
                        \textbf{29}, 47 (1979).
 \bibitem{Farkas}       Z. Farkas, G. Bartels, T. Unger and D.E. Wolf, Phys.
                        Rev. Lett. \textbf{90}, 248302 (2003).
 \bibitem{Geng}         J. Geng, E. Longhi, R.P. Behringer and D.W. Howell, Phys. Rev. E
                        \textbf{64}, 060301-1 (2001).
 \bibitem{Gidaspow}     D. Gidaspow, \emph{Multiphase flow and fluidization.
                        Continuum and kinetic theory descriptions} (Academic
		        Press, San Diego, 1994).
 \bibitem{Gregor}       T. Gregor, U. T\"uz\"un and D.M. Heyes, Powder Technol.
                        \textbf{133}, 203 (2003).
 \bibitem{Herrmann}     H.J. Herrmann, Phys. A \textbf{191}, 263 (1992).
 \bibitem{Iwai}         T. Iwai, C.W. Hong and P. Greil, Int. J. Modern Phys. C
                        \textbf{10}, 823 (1999).
 \bibitem{Kondic}       L. Kondic, Phys. Rev. E \textbf{60}(1), 751 (1999).
 \bibitem{Krupp}        H. Krupp, Adv. Colloid Interface Sci. \textbf{1}, 111
                        (1967).
 \bibitem{Kuwabara}     G. Kuwabara and K. Kono, Jpn. J. Appl. Phys. Part~1
                        \textbf{26}, 1230 (1987).
 \bibitem{Lecoq}        O. Lecoq \emph{et al}., Powder Technol. \textbf{133},
                        113 (2003).
 \bibitem{Leszczynski}  J.S. Leszczynski, Granular Matt. \textbf{5}(2), 91
                        (2003).
 \bibitem{Lubachevsky}  B.D. Lubachevsky, J. Comput. Phys. \textbf{94}(2), 255
                        (1991).
 \bibitem{Luding0}      S. Luding \emph{et al}., Phys. Rev. E \textbf{49}, 1634
                        (1994).
 \bibitem{Luding}       S. Luding \emph{et al}., Phys. Rev. E \textbf{50}, 4113
                        (1994).
 \bibitem{Lyczkowski}   R.W. Lyczkowski and J.X. Bouillard, Powder Technol.
                        \textbf{125}(2-3), 217 (2002).
 \bibitem{Matuttis}     H.G. Matuttis, S. Luding and H.J. Herrmann, Powder
                        Technol. \textbf{109}, 278 (2000).
 \bibitem{Maugis}       D. Maugis and H.M. Pollock, Acta Metall. \textbf{32},
                        1323 (1984).
 \bibitem{McNamara}     S. McNamara and W.R. Young, Phys. Fluids~A \textbf{4},
                        496 (1992).
 \bibitem{Oldham}       K.B. Oldham and J. Spanier, \emph{The fractional
                        calculus. Theory and applications of differentiation and
		        integration to arbitrary order} (Academic Press, New
			York 1974).
 \bibitem{Painter}      B. Painter and R.P. Behringer, Phys. Rev. E
                        \textbf{62}(2), 2380 (2000).
 \bibitem{Pourin}       L. Pournin,  Th.M. Liebling and A. Mocellin, Phys. Rev. E
                        \textbf{65}, 011302-1 (2001).
 \bibitem{Press}        W.H. Press, S.A. Teukolsky, W.T. Vetterling and B.P.
                        Flannery, \emph{Numerical recipes in Fortran90: The art
		        of parallel scientific computing} (Cambridge  Univ.
		        Press, Cambridge, 1996).
 \bibitem{Rapaport}     D.C. Rapaport, \emph{The art of molecular dynamics
                        simulation} (Cambridge University Press, Cambridge,
		        1995).
 \bibitem{Rimai}        D.S. Rimai, L.P. DeMejo and R.C. Bowen, in
                        \emph{Fundamentals of Adhesion and Interfaces, 1995},
		        edited by D.S. Rimai, L.P. DeMejo and K.L.  Mittal
		        (VSP BV Utrecht, Netherlands, 1995), p.~1. 
 \bibitem{Rioual}       F. Rioual, A. Valance and D. Bideau, Phys. Rev. E
                        \textbf{62}(2), 2450 (2000).
 \bibitem{Samko}        S.G. Samko, A.A. Kilbas and O.I. Marichev,
                        \emph{Fractional Integrals and Derivatives. Theory
		        and Applications} (Gordon and Breach, Amsterdam, 1993).
 \bibitem{Seville}      J.P.K. Seville, C.D. Willett and P.C. Knight, Powder
                        Technol. \textbf{113}, 261 (2000).
 \bibitem{Schinner}     A. Schinner, Granular Matt. \textbf{2}(1), 35 (1999).
 \bibitem{Walton}       O.R. Walton and R.L. Braun, J. Rheol. \textbf{30}, 949
                        (1986).
 \bibitem{ZhangGhadiri} Z. Zhang and M. Ghadiri, Chem. Eng. Sci.
                        \textbf{57}(17), 3671 (2002).
 \bibitem{Zhang0}       D. Zhang and W.J. Whiten, Powder Technol. \textbf{88},
                        59 (1996).
 \bibitem{Zhang1}       D. Zhang and W.J. Whiten, Powder Technol. \textbf{102},
                        235 (1999).
\end{thebibliography}
\end{document}